\documentclass{aa}  
\usepackage{graphicx}
\usepackage[varg]{txfonts}

\newcommand{\changed}[1]{#1}                  
\newcommand{\secchanged}[1]{#1}                  

\newcommand{\kms}{km\,s$^{-1}$}
\newcommand{\Msun}{$M_{\odot}$}
\newcommand{\Lsun}{$L_{\odot}$}
\newcommand{\teff}{T_{\rm eff}}
\newcommand{\nii}{[\ion{N}{ii}]}
\newcommand{\oiii}{[\ion{O}{iii}]}
\newcommand{\oii}{[\ion{O}{ii}]}

\defcitealias{perinotto.04}{Paper~I}
\defcitealias{schoenetal.05a}{Paper~II}
\defcitealias{schoenetal.05b}{Paper~III}
\defcitealias{schoenetal.07}{Paper~IV}
\defcitealias{schoenetal.10}{Paper~VII}

\begin{document}

\title{The evolution of planetary nebulae}

\subtitle{VIII. True expansion rates and visibility times\thanks{Dedicated to the memory of Volker Weidemann who died on March 14, 2012, at the age of 87 years.}}

\titlerunning{The evolution of planetary nebulae. VIII.}

\author{R. Jacob \and D. Sch\"onberner \and M. Steffen}

\institute{Leibniz-Institut f\"ur Astrophysik Potsdam, An der Sternwarte 16, D-14482 Potsdam, Germany\\
           \email{[deschoenberner, msteffen]@aip.de}}

\date{Received 21 March 2013 / accepted 15 July 2013}



\abstract
{The visibility time of planetary nebulae (PNe) in stellar systems is an essential quantity for estimating the size of a PN population in the context of general population studies. For instance, it enters directly into the PN death rate determination. 
}
{The basic ingredient for determining visibility times is the typical nebular expansion velocity, as a suited average over all PN sizes of a PN population within a certain volume or stellar system. The true expansion speed of the outer nebular edge of a PN is, however, not accessible by spectroscopy -- a difficulty that we surmount by radiation-hydrodynamics modelling.
}
{We first discuss the definition of the PN radius and possible differences between the observable PN radius and its physical counterpart, the position of the leading shock of the nebular shell. We also compare the H$\alpha$ surface-brightness evolution predicted by our radiation-hydrodynamics models with the recent H$\alpha$ surface-brightness radius calibration of \changed{\citet[][PhD thesis, Macquarie University, Australia]{frew.08}} and find excellent agreement. We then carefully investigate the existing spectroscopic data on nebular expansion velocities for a local PN sample with objects up to a distance of 2~kpc with well-defined round/elliptical shapes. We evaluate, by means of our radiation-hydrodynamics models, how these observed expansion velocities must be corrected in order to get the true expansion speed of the outer nebular edge.
}
{We find a mean true expansion velocity of 42 \kms, i.e. nearly twice as high as the commonly adopted value to date. Accordingly, the time for a PN to expand to a radius of, say 0.9 pc, is only 21\,000$\pm$5000 years. This visibility time of a PN holds for all central star masses since a nebula does not become extinct as the central star fades. There is, however, a dependence on metallicity in the sense that the visibility time becomes shorter for lower nebular metal content.  
}
{These statements on the visibility time only hold for volume-limited samples. Extragalactic samples that contain spatially unresolved nebulae are flux limited, and in this case the visibility time directly depends on the limiting magnitude of the survey. To reach a visibility time of 21\,000 years, the survey must reach about 7 magnitudes below the bright cut-off of the planetary nebula luminosity function. With the higher expansion rate of PNe derived here we determined their local death-rate density as $(1.4\pm 0.5){\times} 10^{-12}$ PN\,pc$^{-3}$\,yr$^{-1}$, using the local PN density advocated by \cite{frew.08}.
}

\maketitle

\section{Introduction}\label{intro}

Formation and evolution of PNe has been a topic of active research for many years. Regardless of whether single or binary evolution is invoked, the expansion behaviour of PNe is, next to their shapes, an important property because it tells us much about the physical processes involved and the environment into which they expand.  The expansion behaviour of a PN determines its visibility time, and a realistic mean value is an important requisite for birth- and death-rate determinations and for estimates of the total PN populations from different formation scenarios (single vs. binary stars) in general population synthesis studies such as those of \citet{MoDe.06} or more recently, \cite{MoDe.12}. The mean visibility time is directly proportional to number of PNe expected in a stellar population. Thus the knowledge of the expansion properties of a PN, their evolution with time, and dependence on the parameters of the whole system (nebular and central-star mass, metallicity, etc.) is paramount for determining a meaningful average visibility time of a PNe ensemble. 

While the (angular) size of a PN is a rather well-defined observable quantity (the exceptions are bipolar and irregular objects), the expansion rate is not. The Doppler split of emission lines, caused by the approaching and receding nebular shells, is an indicator of the nebula's expansion. Generally, either the (maximum) separation of emission line peaks, in the case of spatially resolved objects, or the half widths of emission lines from unresolved objects are used. In both cases it is unclear whether these values represent the true expansion of PNe because density and velocity gradients are not considered. Even more important is the fact that the true expansion velocity, i.e. that of the nebula's outer edge (shock or ionisation front), cannot be measured spectroscopically at all. Nevertheless, this kind of velocity  information that is collected in catalogues, such as that of \citet[][]{W.89}, is extensively used in the literature for statistical studies without any critical reflection.
  
Despite the importance of reliable knowledge about the expansion rates and ages of PNe, progress was not made until it became possible to model formation and evolution of a PN fully self-consistently; i.e., given the mass-loss/wind history at the tip of the asymptotic giant branch (AGB), the combined evolution of stellar remnant and post-AGB wind is simultaneously modelled by means of radiation-hydrodynamics \citep[see, e.g.,][ henceforth Paper~I, and references therein]{perinotto.04}.  Although these models are (geometrically) simple, they are physically sophisticated and close enough to reality to be used to interpret observed line profiles in terms of the general expansion behaviour.   

These very detailed simulations clearly demonstrate that the paradigm of interacting winds \citep[invented by][]{kwoketal.78} in which a powerful stellar wind transfers momentum to the circumstellar matter, thereby creating and accelerating a snow-ploughed shell, is too simplified because the dynamical impact of ionisation by the stellar radiation field is not considered. Using a realistically chosen evolution of the stellar wind \citepalias[see][ for details]{perinotto.04}, it follows that the large pressure gradient set up by ionisation generates and drives a shock into the ambient medium, and this shock henceforth defines the outer PN boundary, $R_{\rm out}$, from the physical point of view. Ultimately, the fast wind of the central star  does not collide directly with the former AGB-wind, since the density and velocity structure of the latter have been completely reshaped by the passage of an ionisation and shock front. Instead, the wind impact is transported to the ionised shell via a ``bubble'' of shock-heated wind material, separated from the shell by a contact discontinuity (or conduction front).

Observationally, the outer edge of a PN may be different from the shock's position, depending on the degree of ionisation and the spectral line used. Once the ionisation front has overtaken the shock, the PN is said to be optically thin to ionising radiation, or more precisely, to Lyman continuum photons.
 
From what is said above it is not too surprising that the ionisation-created shock propagates outwards \textit{independently} of current stellar wind properties. During the beginning when the nebula is still optically thick, the speed is ruled by upstream/downstream gas pressure ratio, but later, during the optically thin stage, the shock's speed depends solely on the electron temperature behind and on the density gradient ahead of the shock (\citealp{FTTB90}; \citealp{Ch97}; \citealp{shuetal.02}; \citealp[][ henceforth Paper II]{schoenetal.05a}). Instead of wind power, it is the stellar radiation field that controls the shock propagation: High stellar temperatures favour a high expansion velocity, maybe temporarily interrupted if a luminosity drop leads to partial recombination, hence lower electron temperatures.  Also, the steeper the upstream density slope, the faster the   shock speed, independently of the density itself.  

The stellar wind is of some importance in so far as it hinders the high-pressure ionised gas to fall back to the stellar surface, simply because the thin but very hot bubble gas has a comparable pressure.   Later, during evolution with time as the wind becomes stronger, the bubble starts to expand and compresses the inner parts of the ionised nebular shell into a dense and optically bright ``rim'' whose expansion (more precisely the propagation of its leading shock) is, however, normally slower than that of the outer shock \citepalias{schoenetal.05a}.  This dense rim appears to be the most prominent part of a PN image (and of its emission line profiles), although it contains only 10 to 20\,\% of the total ionised shell mass.

The situation is different for those PNe that harbour central stars with hydrogen-deficient surfaces (spectral types [WC], [WO]/OVI, PG 1159). Here we have much stronger stellar winds, and wind interaction plays a much more important role in the PN dynamics.  Our models cannot be used for these objects since the history of formation and evolution of PNe with hydrogen-deficient nuclei is still not clear.  

It is interesting to note that metal-poor PNe, which are expected to have weaker winds, expand equally fast, if not faster, than Galactic disk PNe \citep[see][]{jacobetal.12}.  This behaviour was predicted by \citet[][ hereafter Paper VII]{schoenetal.10} and is caused by higher electron temperatures. We also expect that Galactic halo PNe have shorter visibility times than their Galactic disk counterparts due to their faster expansion.
 
In closing these introductory considerations we summarise that a typical PN, regardless of whether being the result of single or binary star evolution, is a dynamically active system consisting of shock waves which react in response to time variable boundary conditions which change drastically during the whole system's lifetime. Even during the final phase of evolution when the central star becomes a hot white dwarf, re-ionisation sets in, and the expanding shock also faces an environment shaped by the previous evolution along the late AGB, viz. by thermal pulses \citep[see, e.g.,][\changed{and Paper~I}]{schoenetal.97}. \emph{We remind that ballistic (or uniform expansion), often directly or implicitly assumed, and shock propagation are mutually exclusive.}
 
In the present work we try to estimate, from appropriately measured ``expansion" velocities, the true expansion to be used in statistical studies as accurately as possible. In doing so, we retrieved expansion velocity measurements from the literature and supplemented them with own recent data (Sect.~\ref{sample.velocity}). We then analysed all these measurements by means of our radiation-hydrodynamics simulations in order to find the necessary correction for determining the real expansion velocity, viz. that of the (outer) PN edge (Sect.~\ref{compar.models}).  As expected, we come up with considerably higher expansion velocities than the hitherto used canonical values of 20--25 \kms.  We demonstrate that a mean expansion velocity of about 40 \kms\ is more appropriate for statistical studies (Sect.~\ref{eval.veloc}). The determination of a typical PN visibility time is given in Sect.~\ref{vis.time}. A closing discussion follows in Sect.~\ref{discuss}.

\section{Sample selection and velocity criteria}\label{sample.velocity}

\subsection{The sample}\label{sample}

We based our investigations on a sample of local PNe compiled by David Frew in his dissertation \citep{frew.08}.  This sample contains 55 objects within a distance of 1 kpc (the \emph{local} sample\changed{)}, and slightly more than 200 within about 2 kpc, with nebular radii up to about 1~pc. Distances (and radii) are based in parts on his new calibration using the H$\alpha$ mean surface brightness (\citealt{frew.parker.06}; \citealt{frew.08}, Chapt. 7). This sample is supposed to be complete only to distances of 1 kpc, but we used the larger 2-kpc sample for better statistical relevance since only for less of about half of the objects useful velocity information is available.  All the necessary information on the objects considered here, if not otherwise said, is entirely based on data presented in Tables 9.4, 9.5, and 9.6 of \citet{frew.08}.

\subsection{Velocity criteria}\label{velocity}

Because the mean visibility time plays such a crucial role in estimating the total PN population, the (true) expansion velocities of PNe must be determined as precisely as possible. This is not done by simply averaging over available spectroscopic velocities, neither from the literature nor new measurements (e.g., those of \citealt{frew.08}) because these data are too inhomogeneous and do not reflect true expansion velocities, as we specify below in detail. Also, we restrict our discussion to objects with more or less closed shells since only then a meaningful correction to  observed Doppler velocities can be found by means of spherically symmetric radiation-hydrodynamics simulations. 

In principle, one has to deal with four different velocities of an expanding nebular shell:
\begin{enumerate}
  \item  the propagation of the (outer) shock whose distance from the star 
         defines the actual PN radius, $R_{\rm out}$, and $\dot{R}_{\rm out}$ 
         is the \emph{true} expansion velocity of a PN;
  \item  the post-shock velocity;
  \item  a representative velocity derived from the peak separation of 
         Doppler split emission lines, provided the spatial resolution 
         is sufficiently high;
  \item  a reprentative velocity from the half width of emission lines of spatially
         unresolved objects.
\end{enumerate}   

\changed{$\dot{R}_{\rm out}$ determines the PN visibility time but cannot be measured spectroscopically. Recent attempts to measure the post-shock velocity (case 2) are promising but need observations of both high spatial and spectral resolution \citep{corradetal.07}. The advantage of the post-shock velocity is that it is close to $\dot{R}_{\rm out}$, with a correction factor that only depends on the shock's Mach number and the adiabatic index of the flow.  For typical PN conditions this correction factor is around 1.2--1.3 (\citealp{Me.04}; \citealp[][ henceforth Paper III]{schoenetal.05b}; \citealp{corradetal.07}). 

Only velocities according to cases 3 and 4 are easily accessible to observations and thus normally listed and also used in the literature.  Both of these velocities are, however, underestimating the nebular expansion and need rather large corrections because the nebular shells with the highest emission measure are rarely those that expand fastest.  For instance, the line peaks correspond to the inner, bright rim of the PN, which expands usually much slower than the fainter, outer nebular regions (called ``shell", or sometimes ``attached shell"). The velocity difference can be as high as about 20 \kms! This general expansion property of PNe is discussed in detail in \citetalias{schoenetal.05a} and in \citet{jacobetal.12} and is due to the combined dynamical effect of ionisation and wind interaction. However, the maximum peak separation is well defined in both the observations \emph{and} the models, allowing us to find reasonable correction factors. These, however, depend on the evolutionary phase and can be as high as about 5, as we will see below. 

In the case of lines from spatially unresolved objects (case~4) it is difficult, if not impossible, to identify a nebular region which corresponds to the half-width velocity.  But also the half-width velocity always underestimates the true nebular expansion velocity by similar amount as for the case of well split emission lines \citepalias[see discussion in][ and Fig. 39 therein]{schoenetal.10}.

Since post-shock velocities are known for rather few objects, we have to resort in this study mainly to case~3 and we selected accordingly} from the Frew 2-kpc sample those PNe for which velocity information from (Doppler) split emission lines is available for the strong [\ion{N}{ii}] and/or [\ion{O}{iii}] lines.  Hydrogen or helium were not used because of their relatively large thermal line widths.  The velocity information is largely based on the compilation of \citet{W.89}, together with a few newer sources. Altogether we selected \changed{78 objects} with radius $<$1.2 pc (from a total of 210) with useful velocity information.  
   
Based on this sample, we discarded a few objects which were classified as strongly bipolar and irregular (5 objects).   We also removed those PNe which are known to have a hydrogen-deficient central star (13 PNe) because their evolutionary history is still unknown.  Nevertheless, they are collected in Table~\ref{Wolf-Rayet} together with other relevant data for comparisons with our main sample.  We also separated those objects which are known to harbour a close binary central object (\changed{6 objects}, Table~\ref{binaries}), \changed{i.e. which have a confirmed period. Objects that experience(d) a common envelope evolution cannot be fully addressed by our 1D-code.} Finally, we ended up with a PN sample consisting of \changed{59 objects} with reasonably known line-peak separations which are listed in Table \ref{basic.data}. 

\begin{figure}
\vskip-2mm
\hskip-3mm
\includegraphics[width=\linewidth]{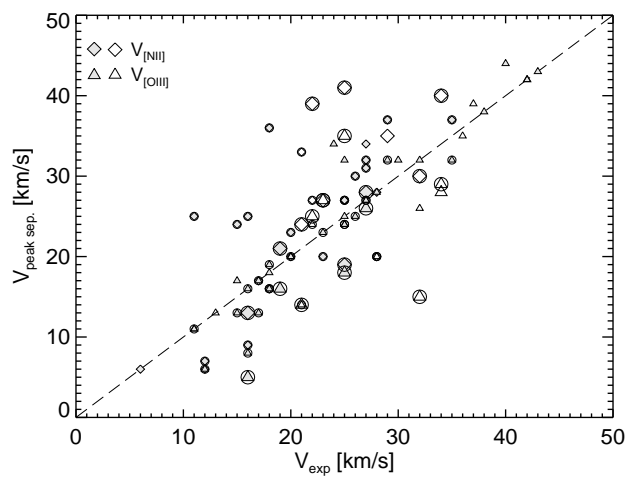}
\caption{Velocities derived from the peak separation of [\ion{N}{ii}] (diamonds)
         and [\ion{O}{iii}] lines (triangles)
         against $V_{\rm exp}$, the  expansion velocities given by \citet{frew.08}, Tables 3.8 
         \changed{(large symbols) and 9.4 (small symbols)}.
         The dashed line is the 1:1 relation. Circled symbols indicate objects for which
         peak separation velocities from both ions are available.  Open symbols mark
         objects from the local sample with distance $\le$1 kpc, filled ones PNe with distances
         $>$1~kpc.     
        }  
\label{vel.frew}
\end{figure}

It is interesting to compare the velocities used here with those provided by \citet{frew.08}.  The latter data set is quite inhomogeneous and consists of own measurements of half widths of spatially unresolved lines of H$\alpha$ (or [\ion{O}{iii}] 5007 \AA\ in some cases), supplemented by literature values (the same that we also use here if no more recent determinations are available). \changed{If both emission lines were available Frew used their mean as value for $V_{\rm exp}$.} The result is shown in Fig.~\ref{vel.frew} where we have plotted velocities, based on peak separations of [\ion{N}{ii}] and [\ion{O}{iii}] lines and listed in Table \ref{basic.data}, against Frew's $V_{\rm exp}$ velocity determinations based on \changed{either} line half widths (his Table 3.8) \changed{or averages of literature values (his Table 9.4)}.\footnote{Throughout the entire paper we denote velocities adopted and listed by \citet{frew.08} with $V_{\rm exp}$, regardless of the sources and methodes they are based on.} 

As expected, the correlation of the peak-separation velocities with \changed{half-width velocities (large symbols)} is really poor, rendering the use of half-width velocities from spatially unresolved lines as quite dangereous:
\begin{itemize}
  \item  for given $V_{\rm exp}$, $V_{\rm \nii}$ is often larger and $V_{\rm \oiii}$
         smaller;
  \item  the difference between $V_{\rm \nii}$ and $V_{\rm \oiii}$, and between both and
         $V_{\rm exp}$, is often about 10 \kms, but can be as large as 15 \kms\ in few cases.
\end{itemize}      

\changed{Furthermore, plain averaging of velocities deduced from different emission lines is also not recommendable since it does not take into account that these lines may originate in different domains within a more complex velocity field, as can be seen from the frequent deviation of the rest of the sample (small symbols) from the 1:1 correlation.}

This figure clearly demonstrates the sort of confusion that exists today about the expansion behaviour of PNe and stresses the importance of a better interpretation of the many high-quality data already available.

\section{Comparisons with hydrodynamical models}\label{compar.models}

\subsection{The hydrodynamical models}\label{models}

The basis of our study is the comparison of theoretical line profiles computed from radiation-hydrodynamic models with the observations, with the final goal to determine true expansion velocities of PNe and to derive relations between nebular expansion and size, if they exist. For this purpose we employ the set of hydrodynamical sequences introduced in detail in \citetalias{perinotto.04} and later supplemented by additional sequences in \citetalias{schoenetal.05a}, \citetalias{schoenetal.05b}, and \citetalias{schoenetal.10}. For a description of all relevant details of our 1D radiation-hydrodynamics simulations the reader is referred to these publications. We note here only that, once the initial circumstellar density distribution, its metallicity, the stellar remnant mass (i.e. the central star mass), and the time-dependent stellar wind are specified, the evolution is followed fully consistently with the changing boundary conditions. Also, all ionisation and recombination processes for the considered elements, together with heating and cooling of the gas, are treated fully time-dependently for the 9 elements listed in Table \ref{tab.element}. 

\begin{figure}
\vskip-2mm
\hskip-3mm
\includegraphics[width=1.\linewidth]{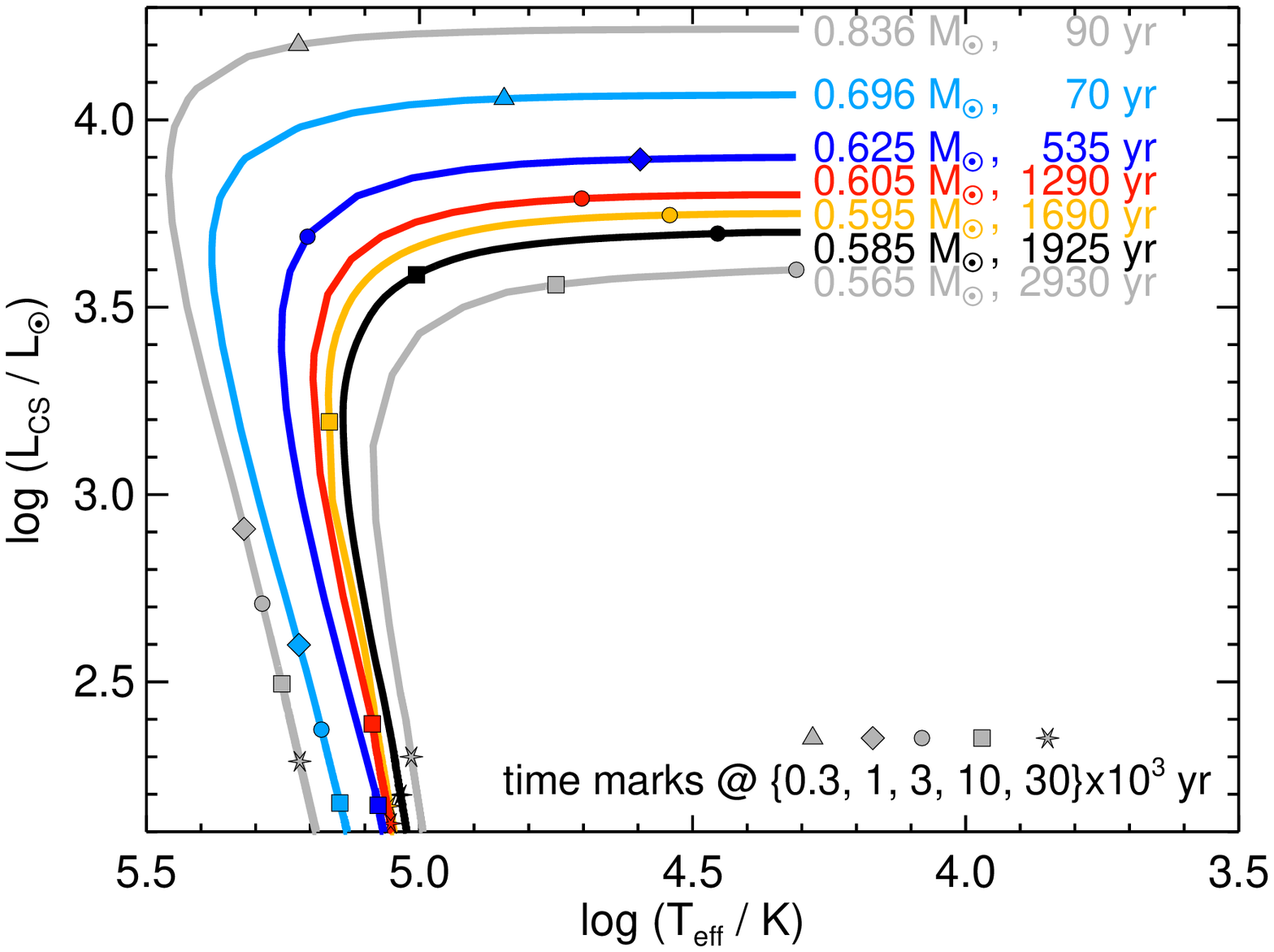}
\caption{\label{HRD}
         Hertzsprung-Russell diagram of post-AGB stellar models used in nebular simulations.
         They are from \cite{schoen.83} and \citet{SB.93} and interpolated (0.595 and
         0.585 \Msun).  The progress of stellar evolution along each track is indicated by 
         time marks at $0.3{\times}10^3$, $1{\times}10^3$,
         $3{\times}10^3$, $10{\times}10^3$, and $30{\times}10^3$ years.  The ``transition time'' 
         from the tip of the AGB to ${\teff =20\,000}$~K is also given for each model sequence 
         and is fully considered for the positioning of the time marks.  
         The colour coding of the individual post-AGB tracks is kept the same throughout 
         the whole paper.   
        }
\end{figure}  
   
\begin{table}[t]           
\caption{Chemical abundance distribution typical for Galactic disk PNe.          
        }
\label{tab.element}
\centering
\tabcolsep=4.5pt
\begin{tabular}{ccccccccc}
\hline
\hline\noalign{\smallskip}
 H     &  He    &  C     &  N    &  O    &  Ne   &  S    &  Cl   &  Ar    \\[1.5pt]
\hline\noalign{\smallskip}
 12.00 &  11.04 &  8.89  &  8.39 &  8.65 &  8.01 &  7.04 &  5.32 &  6.46  \\[1.5pt]
\hline
\end{tabular}
\tablefoot{The abundances, $\epsilon_i$, are given in the usual manner as (logarithmic)
           number fractions relative to hydrogen, i.e. $\epsilon_i=\log\,(n_i/n_{\rm H})+12$.
          }
\end{table}
  
Finally, we have to specify the chemical composition of the circumstellar envelopes since the latter influence the nebular expansion properties via the electron temperature. Our reference abundance distribution for the Galactic disk PNe, $Z_{\rm GD}$, is listed in Table \ref{tab.element}. These abundances are the same as those used in our previous hydrodynamical computations and are, apart from carbon and nitrogen, very close to the most recent solar values \citep[see, e.g.,][]{asplund.09}.   If not otherwise specified, the hydrodynamical models (central-star wind and nebula) presented here have the $Z_{\rm GD}$ elemental composition shown in Table \ref{tab.element}.  
     
\begin{figure*}
\includegraphics[width=0.32\linewidth]{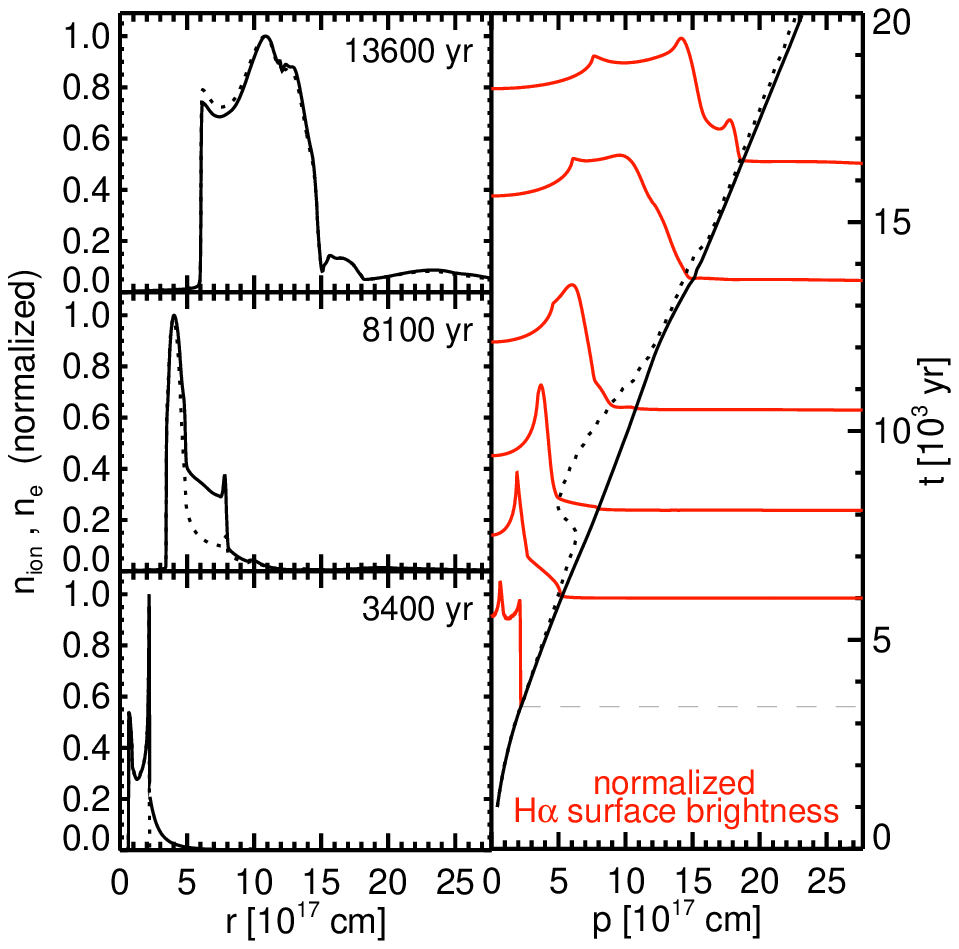}
\hfill
\includegraphics[width=0.32\linewidth]{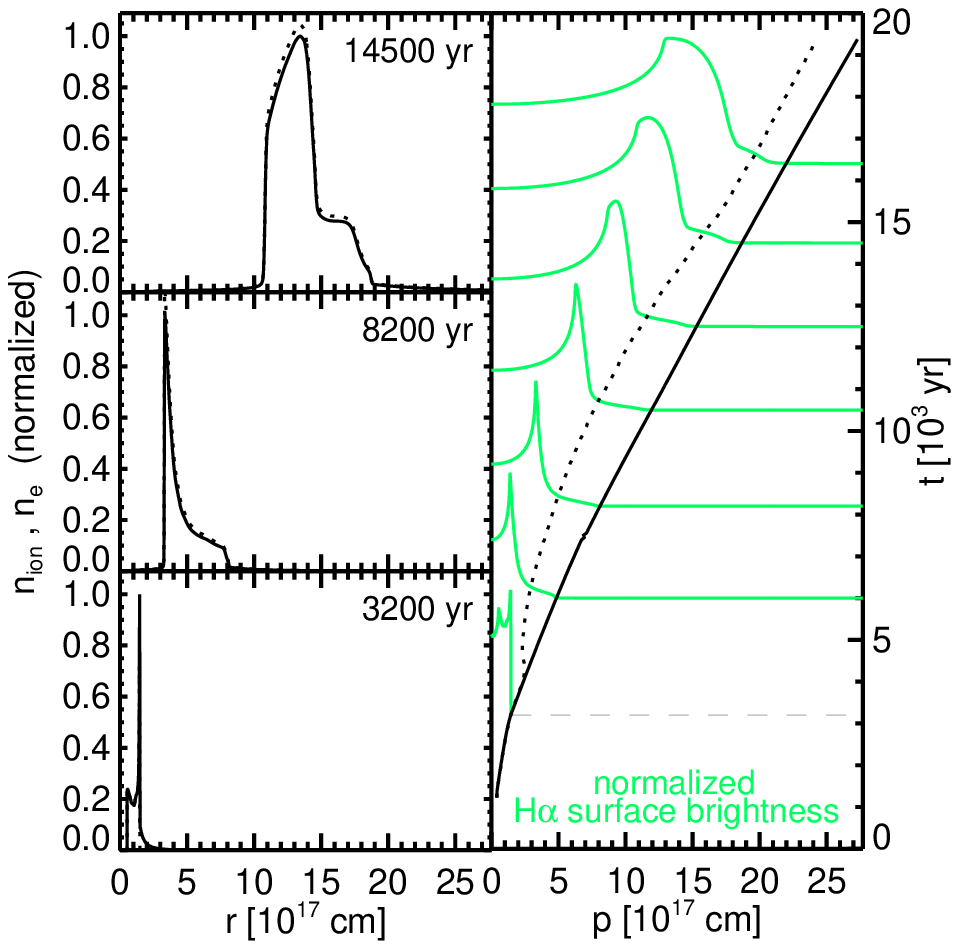}
\hfill
\includegraphics[width=0.32\linewidth]{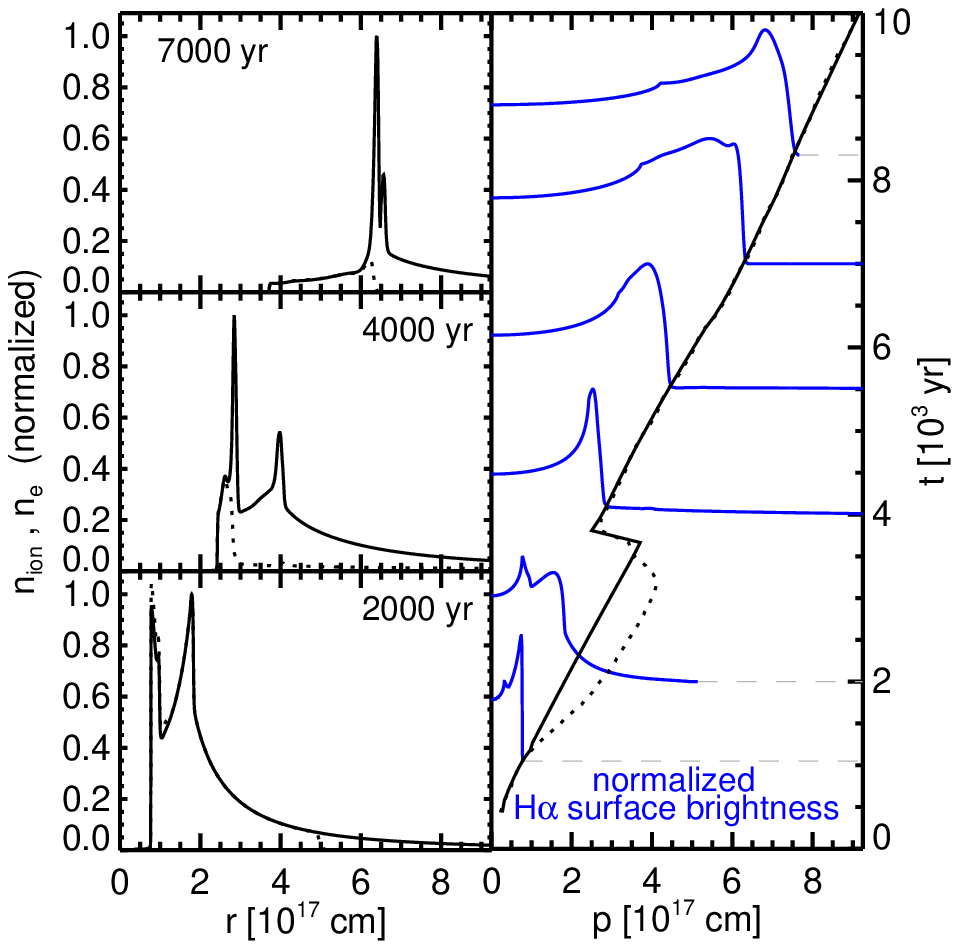}
\caption{\label{r10.vs.rout}  Snapshots of
         model density structures and H$\alpha$ intensity distributions for 3 
         hydrodynamical sequences: 0.605~\Msun\ ``hydro'', (\emph{left}), 0.595 \Msun\ 
         ``${\alpha=3}$'', (\emph{middle}), and 0.625 \Msun\ ``${\alpha =2}$'' (\emph{right})
          at selected times (right ordinate).
         \emph{Left part of the panels}: normalised total ion (solid) and electron densities 
         (dotted) over radius $r$ at the indicated times, for comparisons.
         \emph{Right part of the panels}: evolution of $R_{\rm out}$ (solid) and
         $R_{\rm 10}$ (dotted) with time $t$. At a number of times, defined by the cut of the
         intensity line with the right ordinate, the (normalised) 
         H$\alpha$ intensity profiles are plotted over the impact parameter $p$.
         Note that the extremely sharp density peaks seen in the
         \emph{right panel} at times of 4000 and 7000 years have no correspondence
         in the intensity profiles because the gas is not ionised.
        }
\end{figure*}  
     
Our model sequences try to cover the most relevant combinations between initial envelope density distributions (based on hydrodynamical simulations along the tip AGB, ``hydro'', or simple power-laws, ${\rho \propto r^{-\alpha}}$, with ${\alpha = 2,\, 2.5,\, 3}$) and AGB wind speeds (10 to 15 \kms), and post-AGB masses (0.585 to 0.696 \Msun). The most important ingredient is, however, the stellar mass because the evolutionary speed across the Hertzsprung-Russell diagram \changed{(HRD)} and down to the domain of the white dwarfs are highly dependent on that mass.  We employed the post-AGB model sequences of \cite{schoen.83} and \citet{B.95} together with two interpolated sequences (0.595 and 0.585~\Msun) and show their \changed{HRD} in Fig.~\ref{HRD}. 
 
One immediately sees how sensitive the time evolution depends on central-star mass. Assuming a total nebular lifetime of, say, 20\,000 years one can state that PNe spend most of their life harbouring a very faint, white-dwarf like central star. The exception are those around the lightest nuclei with $\la$0.57 \Msun. The stellar evolution has a profound influence on the surrounding circumstellar AGB material: First rapid ionisation and heating, later when the star fades inward progressing recombination, with some consequences for the radius definition/determination, as we discuss further below.   
   
Before continuing, we want to discuss the so-called ``transition time'' from the tip of the AGB towards, say, 20\,000 K stellar effective temperature.  The transition times depend, like the following evolution, severely on remnant mass.  The values used here (Fig.~\ref{HRD}) are from \cite{schoen.83} and \citet[][ Fig. 5 therein]{SB.93} and are based on the assumption that the zero-point of post-AGB evolution is very close to the AGB, i.e. between 5500~K (0.565 \Msun) and 7000 K (0.696 \Msun).  The transition time is, however, highly dependent on the choice of the zero point of post-AGB evolution and on the post-AGB mass loss in the vicinity of the AGB and must be considered rather uncertain. They cannot, however, be comparable to a typical PNe lifetime because then the density of circumstellar material would become lower than necessary for the formation of PN structures like rims \emph{and} shells.
    
Because of the transition time the formation of a PN occurs at some distance from the stellar surface, i.e. the initial nebular radius is expected to be larger than zero. This fact, however, poses no problem for further considerations, even if our estimates were grossly wrong: Assuming that the AGB envelope continues to expand with 10 (15) \kms, i.e. with no significant acceleration by the stellar wind, initial PNe radii may vary between 0.001 (0.0015)~pc for  0.836~\Msun\ and 0.03 (0.045) pc for 0.565 \Msun. For the typical case of 0.595 \changed{(0.605)} \Msun, initial radii of about 0.015 (0.02) pc follow. Altogether, we can say that any offset from zero nebular radius caused by the transition from the tip of the AGB to the PN-forming region in the HRD is very small compared to PN radii of up to about 1 pc as considered here and can safely be neglected.
     
A set of tools allows us to compute observable quantities from these models, viz. line fluxes, surface brightnesses, and line profiles, either through a central aperture or integrated over the whole model.  These tools are necessary in order to link the model properties to the observations and to check our assumptions on initial conditions, wind evolution, and selection of central star mass. In the past we have compared the predictions of our models with observed properties of well-known round/elliptical PNe in various publications and have found gratifying agreements, at least qualitatively \citep[cf.][]{St.Sch.06}. We have also found no indications that our transition times are grossly wrong \citepalias[see, e.g., Fig. 10 in][]{schoenetal.05b}.

\subsection{Definition of the nebular radius}\label{radius}

In the present paper we connect our models to the PNe sample of \cite{frew.08} not only via emission line splittings but also by using nebular radii. The (true) radius of a model PN, $R_{\rm out}$ is defined by the position of the outer expanding shock, which was set up by ionisation. The observed radius is, however, defined by a suited cut-off value of the surface brightness distribution whose radial position may not agree with that of the shock, especially during phases when the shell matter behind the shock is recombining and its emission becomes weak (see below for more details).   
 
In order to have a homogeneous definition of the nebular radius, \cite{frew.08} declared the H$\alpha$ 10\,\% isophote of a PN image (relative to maximum brightness in H$\alpha$) as the outer edge of this particular object, $R_{\rm PN}$. Since most objects are not round, $R_{\rm PN}$ is a geometric mean of the minor and major semi-axes. In order to be consistent when confronting theory with observation, we decided to proceed in the same way with our models and defined the model's radius, $R_{\rm 10}$, as the radial position in our grid where the H$\alpha$ intensity reaches 10\,\% of its peak value.
   
Since we know the model's radial intensity distribution, we can check the (theoretical) correspondence between $R_{\rm out}$ and $R_{\rm 10}$.  This correspondence is demonstrated in Fig.~\ref{r10.vs.rout} where we compare density distributions with the corresponding surface brightness distributions in H$\alpha$ and both radius definitions.

First of all, we notice that, despite the spherical symmetry of our models, the density and intensity structures are quite complicated, especially for the ``hydro'' sequence. Also, these structures change considerably with time. This is the consequence of ionisation/recombination and wind interaction whose relative importance change during the whole evolution.  
    
More important here is the fact that $R_{\rm 10}$ deviates from $R_{\rm out}$ during \emph{partial} recombination of the shell as seen in the 0.605 \Msun\ ``hydro'' sequence (Fig.~\ref{r10.vs.rout}, left): As long as the model is fully ionised (${t \la 7000}$ yr and  ${t\ga 12\,000}$ yr), the difference between both radii can safely be neglected, the difference being not  more than about 2\,\% (cf. right part of Fig.~\ref{r10.vs.rout}, left panel). But as the central star fades, the outer regions of the model, preferentially the shell, recombine, and their brightness falls quickly \emph{below} the 10\,\% level!  Consequently, $R_{\rm 10}$ becomes significantly smaller than $R_{\rm out}$ (here up to about 40\,\%). Practically, the ``observable'' model radius is now smaller than before for a time span of about 2000 years.  The model PN consists virtually only of an expanding rim.  
    
Recombination lasts only for about 1000 years, followed by reionisation due to continued expansion (and dilution) during which the rim matter becomes accelerated, with the consequence that it swallows later the shell matter and unites thereby both nebular shells (rim and shell) into a single one.  This happens at ${t \approx 12\,000}$ yr, and $R_{\rm 10} \approx R_{\rm out}$ again, with a difference of less than 2\,\% only at the end of our simulation (20\,000 years).
    
The middle panel of   Fig.~\ref{r10.vs.rout} contains the  0.595 \Msun\ ``${\alpha=3}$'' sequence where recombination is unimportant.  Here the shell expands rather fast, leading to small densities with an emission that falls partly \emph{below} the 10\,\% level even without recombination! Thus $R_{\rm 10}$ stays smaller than $R_{\rm out}$ by up to about 20\,\% but keeps up about the same propagation speed.  At the end of our simulation, $R_{\rm 10}$ is smaller than $R_{\rm out}$ by about 10\,\%.
    
The degree of recombination depends sensitively on the nebular density as the star fades. Hence there is the tendency that recombination becomes very important for models with more massive, faster evolving central star.  An example is shown in the right panel of Fig.~\ref{r10.vs.rout} where the model sequence with the 0.625~\Msun\ nucleus is displayed. Here we have the case that during the optically thin period (between 1000 and 3500 years) the halo becomes brighter than the 10\,\% level, and hence $R_{\rm 10}$ is larger than $R_{\rm out}$!   The reason is the still rather high density in the halo because of the initially chosen ``$\alpha = 2$'' density gradient. At about 3800 years, recombination of the shell is \emph{complete} (contrary to the 0.605~\Msun\ case above), and the new outer edge of the model is given by the rim shock, and $R_{\rm 10}$ becomes virtually identical with the position of this shock. After 6000 years the expanding reionisation front of the rim has swallowed the former shell completely, and the former shell's shock continues to propagate with nearly constant speed into the neutral upstream material, engulfing and ionising it. For the rest of evolution, $R_{\rm 10}$ is indistinguishable from $R_{\rm out}$.
    
We note that the PNe does not become ``extinct'' because of nebular recombination due to the fast drop of the stellar luminosity. Rather, we see in the right panel of Fig.~\ref{r10.vs.rout} that the model continues to expand, albeit from a smaller radius (depending on its definition) and reionises, at virtually constant stellar luminosity ($\approx$200\,\Lsun) and temperature ($\ga$100\,000\,K). Generally, nebulae spend most of their lifetime in this stage with low-luminosity central stars.   For the example shown in the right panel of Fig.~\ref{r10.vs.rout} with 0.625~\Msun, recombination starts already at ${R_{10} \approx 4{\times} 10^{17}}$~cm (0.13~pc).   According to the left panel of  Fig.~\ref{r10.vs.rout}, recombination can occur at ${R_{10} \approx 7{\times} 10^{17}}$ cm (0.23~pc) even for 0.605 \Msun, a typical central star mass.
    
Models with slowly evolving low-mass central stars may not recombine at all during stellar fading.  The exact value of $R_{\rm 10}$ relative to $R_{\rm out}$ depends on the density distribution behind  $R_{\rm out}$.  For instance, at the end of our 0.585~\Msun\ simulation, $R_{\rm 10}$ is only 5\,\% smaller than $R_{\rm out}$. Models which recombine and reionise give differences of less than 2\,\% only. Thus we expect that also during the late PN evolution their 10\,\% isophote radius is very close to their real physical radius.  
               
\begin{figure}
\includegraphics[width=1.0\linewidth]{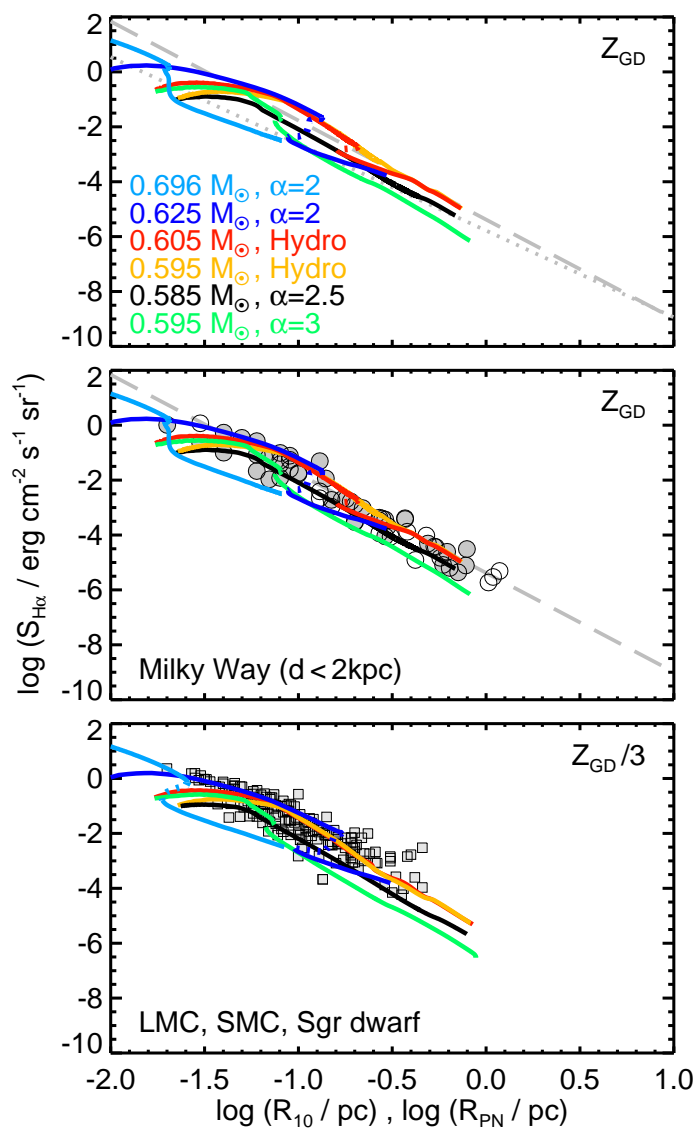}
\caption{The H$\alpha$ surface brightness, $S_{\rm H\alpha}$, as observed and computed 
         from models versus observed/modelled radius.            
         \emph{Top panel}: run of H$\alpha$ surface brightness over $R_{10}$ as predicted 
         by our 1D-radiation-hydrodynamics simulations with various central stars and different
         initial configurations as indicated in the legend. The model sequences are labeled by the
         central-star masses and are described in detail in \citetalias{perinotto.04}
         and \citetalias{schoenetal.05a}. ``Hydro" stands for hydrodynamical modelling
         of dusty AGB wind envelopes along the final AGB evolution, and ``$\alpha$"
         is the exponent of a much simpler power-law density distribution for the
         initial circumstellar neutral shells.  Models are plotted for central stars
         hotter than 20\,000 K until end of simulations. The starting point for the 0.696 \Msun\ 
         sequence is outside the graph at $R_{10} = 0.002$ pc. The non-monotonic behaviour of
         $S_{\rm H\alpha}$ is due to recombination occurring in the models with more 
         massive nuclei.  The dotted parts of some sequences emphasise this brief phase of
         decreasing $R_{10}$. The metallicity of the nebular models, $Z_{\rm GD}$ is given in 
         Table~\ref{tab.element}.The thick dashed line indicates the recent calibration by 
         \citet{frew.parker.06} and \citet[][ Eq. 7.1 therein]{frew.08}, based on 122 carefully 
         selected calibrating sources.  The thick dotted line represents Frew's \citeyearpar{frew.08} 
         calibration of objects with high nebular excitation, defined by the condition that the 
         $\lambda$4686 \AA\ flux is at least 75\,\% of the H$\beta$ flux (Frew's Eq. 7.8).        
         \emph{Middle panel}:  models compared with PNe data from Table \ref{basic.data}.
         The PNe sample is broken down into objects with distances $\le$1 kpc (open symbols) 
         and $>$1~kpc (filled symbols). 
         \emph{Bottom panel}:  the same models as shown in the {\it top panel}, but with 
         down-scaled metallicity $Z_{\rm GD}/3$, and compared to Magellanic Clouds 
         and Sgr dwarf spheroidal galaxy PNe (from Fig. 7.18 in \citealt{frew.08}).
        }
\label{moe3}
\end{figure}

Also more extended model simulations with even more massive (up to 0.940~\Msun, \citetalias{perinotto.04}) and thus much faster evolving central stars demonstrate that a PN will not get ``extinct'' only because the central star fades. The continued expansion leads to more ionisation during the very slow evolution of the central star along the white dwarf cooling sequence. Therefore, the PN visibility of a PN is only limited by the fact that, at some point of evolution, the surface brightness becomes too small and falls below the sky background.\footnote{We assume that the objects in question are spatially resolved.} With the present observational techniques this happens at a surface brightnes of about $10^{-6}$ erg\,cm$^{-2}$\,s$^{-1}$\,sr$^{-1}$ and a radius of about 1 pc \citep[][and see Fig.~\ref{moe3}]{frew.08}. This statement holds for \emph{all} central-star masses!
     
In order to be consistent with the measurements of \cite{frew.08}, we decided to use $R_{\rm 10}$ in all presentations or discussions of our hydrodynamical models where comparisons with observed $R_{\rm PN}$ values are involved.  However, because the 10\,\% radius is not a real physical entity and may vary much with the ionisation state during evolution, as we have seen above, we decided to refer all expansion properties to the physical radius (i.e. to $R_{\rm out}$). This method ensures that our estimates for the real velocities of PNe are independent of the chosen radius definition.
     
Finally, we want to emphasise that a possible difference between $R_{\rm 10}$ and $R_{\rm out}$ is in most cases smaller than the spread beween major and minor axes of elliptical PNe.    
     
\subsection{The H\boldmath$\alpha$\unboldmath~surface brightness--radius relation}\label{SB.radius}

The surface brightness (or intensity) of a model is quite easy to determine and can be used for a comparison with observations, independently of any distance uncertainties. Although our models have only spherical geometry, we assume that they represent, on the average, real objects, which are mostly of elliptical shape, reasonably well.\footnote{The observed number of objects with extreme bipolar or irregular morphology is quite small: 13\,\% \citep{manchetal.00}, 10\,\% \citep{parketal.06, miszetal.08}.} 

The surface brigthness of a PN changes during the evolution not only because of the changing efficiency of converting stellar UV radiation into H$\alpha$ emission and the nebular expansion, but also because the nebula's mass changes as well: mass is steadily swallowed up from the upstream matter by the shock's propagation, increasing thereby the emitting nebula mass.  Modifications may occur during the final drop of the stellar luminosity if recombination prevails and temporarily reduces the ionised nebular mass (see discussion in the previous section).

The predicted changes of the mean H$\alpha$ surface brightness with nebular radius for a number of hydrodynamical sequences is displayed in Fig.~\ref{moe3}, top panel. These simulations cover the relevant mass range from 0.585 \Msun\ to 0.696 \Msun, coupled to various initial neutral envelope configurations \changed{\citepalias[see][or Sect.~\ref{models}]{perinotto.04}.}  The 0.585 \Msun\ sequence has been intruduced in \cite{mendezetal.08}.
   
The sequences are plotted for stellar temperatures higher than 20\,000 K only.  During the early evolution the rapid increase in the ionising flux leads to a simultaneous increase in the ionised and emitting mass, counteracting somewhat the effect of increasing nebular radius: The surface brightness does not change much as compared to the following evolution. The maximum efficiency for converting the stellar continuum UV radiation into H$\beta$ line emission occurs at  $\teff \approx 60\,000$~K, and the models enter a phase of steadily decreasing surface brightness, modified by nebular recombination (if it occurs) which produces the "hooks" caused by the combined effect of decreasing ionised mass and radius ($R_{10}$). The effect of recombination is especially well seen in the 0.625 \Msun\ and 0.696 \Msun\ sequences where it is strongest: There appears a rapid drop of $S_{\rm H\alpha}$ around ${R_{10}\approx 0.1}$~pc and $R_{10}\approx 0.02$ pc, respectively, because of stellar fading, and a decreasing/stalling radius because of recombination.

We have also plotted in Fig.~\ref{moe3} (middle) the objects from Table \ref{basic.data}. These objects form a mixed sample: some are distance calibrators, the others have distances being calibrated, which explains the dispersion seen in the plot around Frew's H$\alpha$ surface brightness--radius calibration (his Eq. 7.1) which is shown in the top and middle panel of Fig.~\ref{moe3} for comparison. \changed{Because of our sample selection process, the dispersion of PNe around the calibrating relation is smaller than found by \citet[][ his Fig.~7.1]{frew.08} for the whole local sample.} The H$\alpha$ surface brightness limit of the sample from Table~\ref{basic.data} is at about $10^{-6}$ erg\,cm$^{-2}$\,s$^{-1}$\,sr$^{-1}$, corresponding to a PN radius of about 1~pc. 

Although our simulations do not extend to the very largest observed radii, especially for the sequences with more massive central stars\footnote{The computational domains of our simulations were optimised for the more interesting phases that occur during the crossing of the HRD and somewhat beyond, and not for the late stage with an extremely slowly evolving hot white dwarf as central object.}, we conclude from Fig.~\ref{moe3} that there exists a very gratifying agreement between the observations and our models, making us confident that the models provide meaningful information also about the expansion properties of real PNe: Our different model sequences exactly cover the whole range of observations, from the smallest to nearly the largest objects.  The following facts are interesting:

\begin{itemize}
  \item  The bright models of the 0.625 \Msun\ sequence nicely ``embrace'' the  brightest
         PNe with $R_{\rm PN} \la 0.1$ pc.  This model property is especially pronounced for
         the extragalactic sample (see below) shown in the bottom panel of Fig.~\ref{moe3}.
  \item  Models along the low central-star luminosity phase of the 0.696 \Msun\ sequence have virtually 
         no observed counterparts: The stellar luminosity is already too low ($<$500 \Lsun) 
         because the models are beyond maximum steller effective temperature before a typical 
         minimum observable radius, 0.02~pc, is reached!
  \item  The models around the nuclei of 0.605, 0.595 (``hydro'' only), and 0.585~\Msun\ 
         fit the observations best.     
  \item  The models of the 0.595~\Msun\ ``$\alpha=3$" sequence reach very high excitations and 
         also maintain them when the central star fades because of their extreme dilution.    
         They develop the lowest surface brightnesses for given radii, and their position in
         the $\log {S}_{\rm H\alpha}$-$ \log R_{10}$ plot corresponds closely to Frew's
         \citeyearpar{frew.08} calibration for high-excitation nebula (his Eq. 7.8, shown in the 
         top panel of Fig.~\ref{moe3}). 
\end{itemize}   
   
The slope of the observed surface brightness--radius relation shown in Fig.~\ref{moe3}, $-3.61(\pm0.11)$, nicely confirms that the nebular mass increases steadily because the shock swallows matter which has been ejected earlier \citep{frew.08}. Moreover, it  consistently provides a third hint, next to PN halo observations \citep[e.g.][]{sandin.08} and interpreting shapes of PN emission line profiles \citepalias{schoenetal.05a}, that the upstream density slope is steeper than $1/r^2$.\footnote{Neglecting changes of nebular morphology in time, already a simple back-on-the-envelope analysis of an expanding ionised shell of constant density shows that $S \propto M^2 R^{-5}$.  Since the observed $S\propto R^{-3.6}$, $M\propto R^{0.7}$ follows. On the other hand, if we assume $\rho \propto R^{-\alpha'}$ with constant $\alpha'$, we have $M\propto R^{3-\alpha'}$. Hence, $\alpha'$ had to be larger than 2, i.e. 2.3 here.} In order to produce those steeper density gradients, the mass-loss rates must steadily increase until the remnant leaves the tip of the AGB since the wind velocities do not change much during the final AGB mass-loss evolution \citep[cf.][]{steffenetal.98, verbetal.11}.
    
A rather steep density gradient of the initial \changed{envelope} structures is thus important for long-time simulation.  Only because the 0.585, 0.595, and 0.605 \Msun\ sequences have ${\alpha > 2}$ \secchanged{for larger radii}, they are able to follow the observed trend of $S_{\rm H\alpha}$ with radius. \secchanged{Linear extrapolation of the sequences around 0.696 and 0.625 \Msun\ with ${\alpha=2}$ indicates that their surface brightnesses decrease too slowly with radius} (see the run of $S_{\rm H\alpha}$ after recombination). Physically speaking, these models \secchanged{are expected to} accrete too much matter \secchanged{during their further expansion}. In reality, we can also expect a steeper circumstellar density profile, i.e. ${\alpha > 2}$, for the \secchanged{later} evolution not simulated by us \secchanged{for these two sequences. The flatter decrease in $S_{\rm H\alpha}$ with radius (also during the early stages of the sequences around the less massive central star) is mainly caused by the ${\alpha \approx 2}$ density profile in the vicinity of the central star. This choice is inevitable since mass-loss rates evolve through a maximum at the tip of the AGB. However, the radii up to which ${\alpha \approx 2}$ holds depend on the timescale this final mass-loss rate can be considered approximately constant. PNe around more massive central star are expected to form closer to their central star due to their shorter transition time. Therefore, ${\alpha = 2}$ seems to be a justified choice for a substantial share of their total lifetime.}

For a very brief period the shock may run through density regions modified by a thermal pulse during the previous AGB evolution, like in the case of the 0.605~\Msun\ ``hydro'' simulation reported below in Sect.~\ref{exp.prop.mod}.   The corresponding variation of the shock propagation speed is only temporary and of no real significance for the average expansion behaviour.
   
The bottom panel of  Fig.~\ref{moe3} compares another set of models with PNe from the Magellanic Clouds and the Sgr dwarf spheroidal galaxy.  Data are taken from \citet[][ see Fig. 7.18 therein]{frew.08}, but a separation into different morphology classes cannot be made. For these PNe samples the radii are known because they belong to stellar populations with well-known distances. The model parameters are the same as in the two upper panels,  but the models are computed with a (scaled-down) metallicity, $Z_{\rm GD}/3$, \changed{for the circumstellar matter. The grid of post-AGB tracks of \cite{schoen.83,SB.93,B.95} does not attribute to different metallicities. The consequences of this aspect of central-star evolution for the nebular evolution is expected to be moderate compared to the uncertain $Z$-dependence of AGB mass-loss history, and has therefore been neglected in the pilot study of \citetalias{schoenetal.10}.}\footnote{\changed{The metallicity-dependent post-AGB models of \citet{vw.94} lack a consistent treatment of the mass loss from the AGB into the post-AGB phase which mainly comes to light in different transition times. The latest Z-dependent post-AGB models of \citet[][Chapt. 8]{kitsikis.07} do model the transition into the post-AGB phase similar to \citet{B.95}. However, due to numerical problems he only provided models with central-star masses up to 0.600\Msun, and thus his model grid is also not suited to our study.}}  The sequences in the bottom panel of  Fig.~\ref{moe3} differ a bit from the other ones, and they extend to larger radii because they expand faster \citepalias[see][]{schoenetal.10}.  We see that all the conclusions made above, including those concerning the 0.625 and 0.696 \Msun\ models, remain valid!
   
Some concluding remarks:  The close correspondences between the different samples on one side, and with our radiation-hydrodynamics models on the other side, seen in Fig.~\ref{moe3} shows that (i) we have, for the first time, a reliable distance calibration for Galactic PNe, and (ii) a reasonable description of long-time PN evolution by radiation-hydrodynamics simulations.  Figure \ref{moe3} suggests that, if one uses the nebula surface brightness to derive the distance to the object, a typical uncertainty of $\pm$0.1 dex (or 25\,\%) is to be expected.  Our model simulations suggests that a considerably part of this dispersion appears to be intrinsic and is due to the mass distribution of the central stars and variations of the initial envelope structures.
   
The run of the models in Fig.~\ref{moe3}, together with the distribution of PNe, suggest a possible refinement of Frew's \citeyearpar{frew.08} calibration by imposing a piecewise linear calibration instead: \changed{below ${\log R_{\rm PN} = -1.0}$ a} flatter straight line, and above a steeper straight line.

\subsection{Measured velocities and their interpretation}\label{vel.mod}

An overview of the expansion properties of our models and the observations listed in Table \ref{basic.data} is shown in the different panels of Fig.~\ref{moe1}. The evolution is either indicated by the size of the nebula/model radius, $R_{\rm PN}$, $R_{10}$, or $R_{\rm out}$ (left panels) or the absolute brightness of the central star, $M_V$ (right panels). The use of the stellar brightness has the advantage that the mass-dependent evolutionary time scales are taken care of. $M_V$ is a nearly mass independent proxy of central star evolution, similar to $T_{\rm eff}$ but with no ambiguity. For all masses the maximum stellar effective temperature corresponds to an absolute magnitude ${M_V}$ between 4.0\ldots 4.5.  Brighter central stars are at the high-luminosity horizontal part of their evolution, while fainter stars are beyond maximum stellar temperature and are descending to the white-dwarf domain (cf. Fig.~\ref{HRD}).

It is advantageous to use the stellar magnitude because the central star is the main driver of PN formation and evolution, and common properties of the models are clearly recognisable.  The disadvantage is that the information of the evolutionary speed is lost.  We thus concluded to use in Fig.~\ref{moe1} both ways of presenting the velocity evolution of PNe.

\subsubsection{Expansion properties of the models}\label{exp.prop.mod}

\begin{figure*}[!t]
\includegraphics[width=1.0\textwidth]{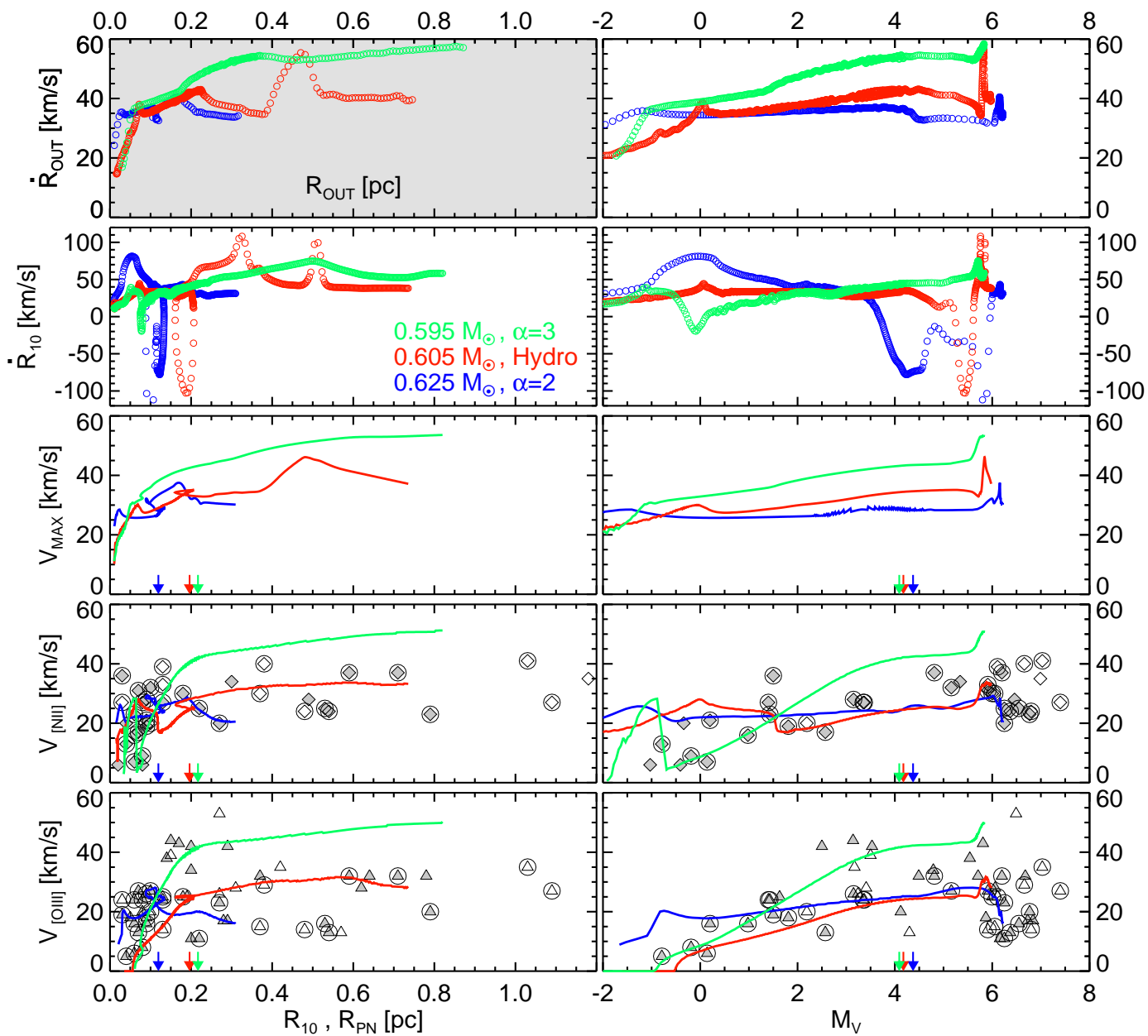}
\caption{Plots of different types of ``expansion'' velocities, observed/modelled, 
         over nebular radius, $R_{\rm PN}$, $R_{10}$, or $R_{\rm out}$ (\emph{left}) and 
         central-star absolute magnitude, $M_V$, of observations/models (\emph{right}).  
         Velocities of the leading edge, $\dot R_{\rm out}$ (\emph{top row})
         and of the H$\alpha$ 10\,\% isophote, $\dot{R}_{10}$ (\emph{second row}) as
         predicted by the models.    Without any loss of
         information, velocities for only three model sequences are plotted, 
         0.595 \Msun\ ``${\alpha=3}$'', 0.605 \Msun\ ``hydro'', and 0.625 \Msun\ 
         ``${\alpha=2}$'', for more clarity.   The top left panel is grey because it is
         the only one with the true model radius, $R_{\rm out}$, as abscissa.
         Note also the extended velocity range
         belonging to the panels in the second row.  The evolution through the loops in
         the left panel, second row occurs clockwise.
         \emph{Third row}:  maximum gas velocities,$V_{\rm max}$,  within the model nebulae 
         from the selected sequences (see text for details).  
         The vertical arrows on the abscissae
         mark the nebular radii or stellar brightness where the central star models reach
         their maximum effective temperatures.  
         These arrows separate between younger models (and objects) with luminous central 
         stars and older ones with faint central stars.   Up to this stage $V_{\rm max}$
         corresponds in all models to the post-shock velocity, $V_{\rm post}$.
         \emph{Fourth and bottom row}: peak-separation velocities
         $ V_{\rm \nii}$ and $V_{\rm \oiii}$, as they follow from central 
         line-of-sight line profiles computed from the models (solid lines) and measured 
         (symbols), as given in Table \ref{basic.data}, respectively.   
         The sample of Table \ref{basic.data} is again boken-down into open symbols ($D\le 1$~kpc)
         and filled symbols ($D > 1$ kpc). Circles around symbols indicate objects
         for which we have velocities from both nitrogen and oxygen ions.
         Zero model velocities mean that the line is not split at an assumed instrumental
         resolution of 10 \kms.
        }
\label{moe1}
\end{figure*}

The first two rows of panels in Fig.~\ref{moe1} show the expansion properties of three representative hydrodynamical model sequences from \citetalias{perinotto.04} and \citetalias{schoenetal.05a}, viz. with 0.595 \Msun\ ``${\alpha=3}$'', 0.605 \Msun\ ``hydro'', and 0.625~\Msun\  ``${\alpha=2}$''. Plotted are the evolution of the models' leading edge, $\dot R_{\rm out}$ (first row of Fig.~\ref{moe1}), and the H$\alpha$ ``10\,\% isophote'' radius, $\dot R_{10}$ (second row of Fig.~\ref{moe1}), over $R_{\rm out}$/$R_{10}$ and stellar absolute magnitude $M_V$.  The total simulation times are about 20\,000 years for the 0.595 and 0.605~\Msun\ cases, but only 10\,000 years for 0.625~\Msun. The model sizes, $R_{10}$ ($R_{\rm out}$), finally achieved are 0.77 (0.87), 0.74 (0.75), and 0.30 (0.32)~pc, and the corresponding mean expansion speeds as defined by final nebular radius divided by simulation time, $\langle\dot R_{10}\rangle$ ($\langle\dot R_{\rm out}\rangle$), are 39.5 (44.5), 36.1 (36.8), and 28.9 (30.6)~\kms, resp.  The other sequences from Fig.~\ref{moe3} are not plotted for clarity because their expansion properties are similar and do not provide additional information.

All aspects of the evolution of  $\dot R_{\rm out}$ are discussed in detail in \citetalias{schoenetal.05a} and  \citetalias{schoenetal.05b}.  We repeat only aspects relevant here. In all cases shown the shock velocity, $\dot R_{\rm out}$, jumps up to $\sim$35 \kms\ during the optically thick stage (independently of stellar mass). The further acceleration during the optically thin phase across the HRD is rather modest and depends on electron temperature and upstream density profile. The shock speed decreases during recombination as the star passes maximum effective temperature, or correspondingly ${M_V \approx 4}$, if recombination occurs (0.605 and 0.625 \Msun).  In the latter case, $\dot R_{\rm out}$ levels off later at about 35 \kms\ since the upstream density slope remains constant (``${\alpha = 2}$''). In the 0.605~\Msun\ case, a brief excursion of high expansion speeds (up to about 55 \kms) occurs because the shock passes a density ``trough'' formed during the last thermal pulse on the AGB with its big mass-loss variations \citep{schoenetal.97, steffenetal.98}. The shock accelerates first during its ``downhill'' propagation, followed then by a corresponding deceleration while climbing ``uphill''.         

Before discussing in detail the behaviour of $\dot R_{10}$ seen in the second row of Fig.~\ref{moe1}, we remind the reader that the H$\alpha$ ``10\,\% isophote'' velocity is, like $\dot R_{\rm out}$, not a matter velocity.  Thus it may even quickly change its direction because the position of $R_{10}$ depends on the density structure of the outer nebula edge and on the ionisation stage as well, as discussed in Sect.~\ref{radius} (see also Fig.~\ref{r10.vs.rout}).  Normally,  along the bright part of evolution (${M_V < 4}$), $\dot R_{10}$ closely follows the shock speed, $\dot R_{\rm out}$, but during recombination/reionisation the situation changes completely:  while $R_{10}$ recedes, $\dot R_{10}\rightarrow \hbox{--100}$ \kms\ for a very brief time!  During reionisation, $R_{10}$ catches up with $R_{\rm out}$.
          
The 0.605 \Msun\ ``hydro'' sequence has a quite  ``complicated'' mass-loss history, thus $\dot R_{10}\rightarrow$ +100 \kms\ during reionisaion (at ${R_{10}\approx 0.32}$~pc). The second velocity ``spike'' at  ${R_{10}\approx 0.51}$~pc is caused by a density ``trough'' which is the signature of the last thermal pulse on the AGB \citep[see][ Fig. 19 therein]{steffenetal.98}.   The high value of $\dot R_{10}$ at ${R_{10} \approx 0.05}$~pc of the 0.625~\Msun\ ``$\alpha=2$'' sequence is caused by the bright halo (cf. Fig.~\ref{r10.vs.rout}, right panel).
     
The third row of panels in Fig.~\ref{moe1}  displays the maximum gas velocities, $V_{\rm max}$, occuring in the models.  As expected, the behaviour of $V_{\rm max}$ is much ``smoother'' than those of $\dot R_{\rm out}$ and $\dot R_{10}$. During the high-luminosity phase of evolution the highest gas velocity occurs immediately behind the shock at $R_{\rm out}$ of the model PN but is, of course, somewhat lower  than the shock speed.  This picture may change during recombination: if the high-speed matter shell recombines too much, only the inner more slowly expanding rim remains ionised and  visible.  This explains the temporary decrease in $V_{\rm max}$ seen for the 0.605 \Msun\ and 0.625 \Msun\ sequences. The broad velocity peak shown by the 0.605 \Msun\ ``hydro'' sequence around ${R_{10} \approx 0.5}$~pc corresponds to the spike explained in the previous paragraph: It is due to shock acceleration during ``downhill'' propagation, followed by a corresponding deceleration while climbing ``uphill''.

\subsubsection{Interpreting the observations}\label{int.obs}

We simulated the line emission and line profiles from each model and measured the peak separation velocities $ V_{\rm \nii}$ and $V_{\rm \oiii}$ for the central line-of-sight.  The results for the three sequences considered are plotted in the two bottom rows of panels of Fig.~\ref{moe1}.  Interesting insights are gained by comparing these Doppler velocities with the corresponding velocities displayed in the two top rows of panels:
\begin{itemize}
  \item  The peak-separation velocities $ V_{\rm \nii}$ and $V_{\rm \oiii}$ behave similarly 
         as $V_{\rm max}$ in so far as they increase, on the average, also with evolution but
         often do not trace the maximum gas velocity.     
  \item  For $V_{\rm \oiii}$ the initial velocity values are 
         far below the initial wind velocities of 10~\kms\ or 15 \kms\ for all three sequences. 
         \emph{Velocities measured from O$^{2+}$ lines suggest obviously an acceleration of nebular
               expansion which is not real.} 
  \item  Except during the short phases of rapid variations of $R_{10}$, \emph{the peak-separation
         velocities, $V_{\rm \nii}$ and $V_{\rm \oiii}$, are always {smaller} than $\dot  R_{10}$}.         
\end{itemize}

The different velocities derived from N$^+$ and O$^{2+}$ for the younger objects (${M_V\la 1}$) is a consequence of ionisation stratification in these models: most of N$^+$ is concentrated in the outer regions with high velocities, while O$^{2+}$ resides in the inner, initially slowly moving nebular regions. Later, when the nebula becomes optically thin, N$^+$ is a minority species and is more evenly distributed throughout the nebula, similarily to O$^{2+}$, and both ions are now tracing the same velocities. The reason for the sudden drop of $V_{\rm \nii}$ seen at ${M_V\approx -1}$ (0.625 and 0.595~\Msun) and ${M_V\approx 1.5}$ (0.605 \Msun) (fourth right panel of Fig.~\ref{moe1})  is the increasing dominance of the only slowly expanding rim matter in the \nii\ line profile after the thick/thin transition and the growing pressure of the wind-blown hot bubble. Velocity differences may also occur at late stages (${M_V\ga 5}$) when recombination establishes again an ionisation stratification.
   
The bottom two rows of panels of Fig.~\ref{moe1} also contain the measured velocities of our locally selected PN sample (Table~\ref{basic.data}).  It is evident that plotting over nebular radius does not appear very conclusive:  Very large nebulae have certainly only a faint, highly evolved central star, but smaller nebulae have either a bright (less evolved) \emph{or} a faint (far evolved) nucleus, depending on its mass. Using the stellar $M_V$ instead separates evolved and unevolved PNe according to the brightness (= evolutionary stage) of the stellar core. We consider a PN around a faint central star (${M_V\ga 5}$) as evolved, irrespective of its size (age). Indeed, the right panels show that our sample contains more evolved than unevolved objects, in (at least qualitative) agreement with the prediction of the theory of post-AGB evolution which predicts a rather fast evolution across the \changed{HRD} within a time span of less than 10\,000 years for AGB-remnants of about 0.6~\Msun\ (cf. Fig.~\ref{HRD}).

The bottom panels of Fig.~\ref{moe1} show an encouraging agreement between theory and observation, although our models do not cover the whole range of nebular sizes. We emphasise that our 0.605~\Msun\ sequence gives a very good description of the observed average late velocity behaviour, both in $V_{\rm \nii}$ and $V_{\rm \oiii}$! The final model peak separation velocities as one would observe from the Doppler split mission lines are about 33~\kms\ for \nii\ and about 28~\kms\ for \oiii. Also, a decrease in the (observed) peak separation velocity with radius (age) is not detectable, in agreement with the prediction of our simulations.
    
\begin{figure}
\vskip-2mm
\hskip-5mm
\includegraphics*[width=1.05\linewidth]{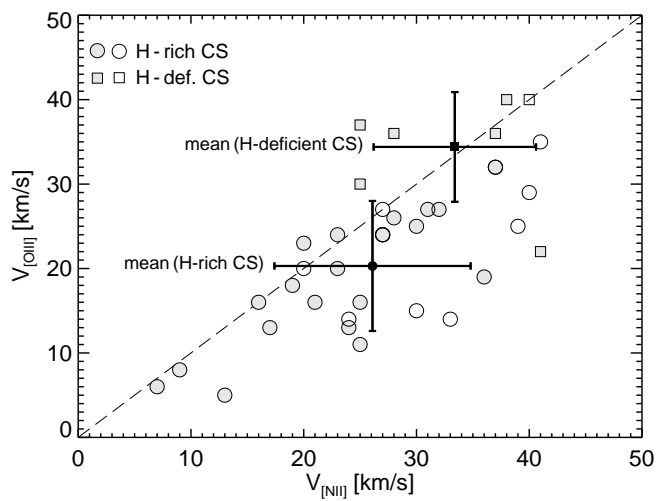}
\caption{\changed{
         Velocities derived from the peak separation of [\ion{O}{iii}] against their corresponding 
         [\ion{N}{ii}] values for PNe with H-rich (circles) and H-deficient (squares) central stars.
         Open symbols mark objects from the local sample with distance $\le$1 kpc, filled ones 
         PNe with distances $>$1~kpc. The mean values and standard deviations for the two subsamples
         are also shown (see text for details).}     
        }  
\label{moe.vels}
\end{figure}

For the PNe sample displayed in Fig.~\ref{moe1}, there are 31 objects (i.e. about 50\,\% of the whole sample considered) with measurements of both peak separation velocities $ V_{\rm \nii}$ and $V_{\rm \oiii}$. \changed{The differences between the velocities deduced from the two emission lines are shown in Fig.~\ref{moe.vels}. We can perform simple statistical exercises:}
\begin{itemize}
  \item  Their averages are
         \begin{eqnarray}
                \langle  V_{\rm \nii}\rangle & = & 26.1 \pm 8.7 \ \ \mbox{\kms}, \nonumber \\
                \langle V_{\rm \oiii}\rangle & = & 20.3 \pm 7.7 \ \ \mbox{\kms}, \nonumber
         \end{eqnarray}
         indicating a  higher average value from N$^+$ lines (by nearly 30\,\%).
         The uncertainties given here and in the following are the 1$\sigma$ width of 
         the respective velocity distributions. 
\changed{
         Interestingly, the corresponding values for the 7 objects with hydrogen-deficient stars
         are, on the average, higher, especially those from O$^{2+}$:
         \begin{eqnarray}
                \langle  V_{\rm \nii}\rangle & = & 33.4 \pm 7.2 \ \ \mbox{\kms},  \nonumber \\
                \langle V_{\rm \oiii}\rangle & = & 34.4 \pm 6.5 \ \ \mbox{\kms}.  \nonumber   
         \end{eqnarray}     
}
  \item  Averaging only over large nebulae ($R_{\rm out}\ge 0.2$ pc, 12 PNe, hydrogen-rich only) 
         does not change the picture:  
         \begin{eqnarray}
                \langle  V_{\rm \nii}\rangle & = & 29.4 \pm 7.4 \ \ \mbox{\kms},  \nonumber \\
                \langle V_{\rm \oiii}\rangle & = & 22.3 \pm 8.5\ \ \mbox{\kms}.  \nonumber
         \end{eqnarray}
         Only one object with a hydrogen-deficient nucleus falls in this category, i.e. 
         \object{NGC 5189} with $V_{\rm \nii} = 25$~\kms\ and 
         $V_{\rm \oiii} = 37$~\kms\ (Table~\ref{Wolf-Rayet}). 
  \item  A similar result follows for objects with faint central stars (${M_V \ge 5}$, 16 PNe),
         many of which are also more massive ones:
         \begin{eqnarray}
                \langle  V_{\rm \nii}\rangle & = & 30.1 \pm 6.6 \ \ \mbox{\kms},  \nonumber \\
                \langle V_{\rm \oiii}\rangle & = & 22.1 \pm 7.4 \ \ \mbox{\kms}.  \nonumber
         \end{eqnarray}
         Also here we have only one PN with a hydrogen-deficient central star, \object{JnEr 1},
         with  $V_{\rm \nii} = 41$~\kms\ and 
         $V_{\rm \oiii} = 22$~\kms\ (Table~\ref{Wolf-Rayet}).    
\end{itemize}

Altogether one concludes from this excercise (and also just by visual inspection of \changed{Fig.~\ref{moe.vels}}) that velocities derived from N$^+$ lines have the tendency to be higher than those based on O$^{2+}$ lines, \changed{as also demonstrated by the analysis of \cite{MPMS06}.}  If we average over the 31 velocity differences of individual objects,  $\langle V_{\rm \nii} - V_{\rm \oiii}\rangle = 5.8\pm 5.6$ \kms. We note that the rather large dispersion of this mean difference between the nitrogen and oxygen velocities is intrinsic and originates in the nebular evolution during which this difference varies considerably, as becomes evident in the next section.

For the 7 objects with hydrogen-deficient nuclei these differences are predominantly negative (cf.~Table~\ref{Wolf-Rayet}), but ${\langle V_{\rm \nii} - V_{\rm \oiii}\rangle = -1.0\pm 9.9}$ \kms\ only because of \object{JnEr\,1} with ${V_{\rm \nii} - V_{\rm \oiii} = 19}$ \kms. The central star of this particular object is of the PG\,1159 spectral type and very faint, ${M_V=6.8}$ mag.  The nebula is expected to reionise, with a positive velocity gradient combined with ionisation stratification, which then would explain that ${V_{\rm \nii}/ V_{\rm \oiii} > 1}$. The other 6 objects have more luminous central stars (${M_V < 3.6}$) with a mean velocity difference  ${\langle V_{\rm \nii} - V_{\rm \oiii}\rangle = -4.3\pm 5.0}$ \kms. A ratio $V_{\rm \nii}/ V_{\rm \oiii} < 1$ indicates a negative radial velocity gradient within their nebula.  From the hydrodynamical point of view, such a negative velocity gradient is the signature of strong interaction with the stellar wind which forces the inner nebular regions to expand faster than the (outer) shock set up earlier by ionisation. 
     
Negative values of $V_{\rm \nii} - V_{\rm \oiii}$ are not found in our sample of 31 PNe with normal, i.e. hydrogen-rich, central stars (except \object{NGC 2438} with --3 \kms, if taken at face value). Obviously, nebulae with normal central stars are dominated by a positive radial velocity gradient, in contrast with what we have learned above from nebulae around hydrogen-deficient nuclei.  This is only possible because interaction with the stellar wind remains less important than photoionisation in determining the hydrodynamics of these PNe. These two obviously different modes of post-AGB evolution should always be distinguished.

\changed{A possible contribution to the line profiles that is not modelled by our 1D-code is turbulence. Its additional pressure component may in principle introduce locally small alterations to the kinematics of the model PNe. However, estimating the true contribution of turbulent motions within PNe from observations is non-trivial \citep[cf.][and refs. therein]{sabbadin.08}. 

The monotonic radial increase in the velocity towards the outer shock within the shell, as it is clearly evident from reconstruction of observations in different emission lines of PNe with still persisting ionisation stratification, and as it is interpreted as the pattern of a rarefaction wave behind the D-type ionisation front, rules out substantial turbulent motions in the shell. However, the wind-compressed rim may be the more prone to turbulence the stronger the central-star wind, which is expected to be met in PNe around H-deficient central stars \citep[e.g.][]{Me.02}. \citet{gesicki.06} also estimated that only about 1/10 of PNe around H-rich central stars needed the assumption of a turbulent velocity component in order to explain their emission line profiles whereas the share for PNe around H-deficient central stars is rather 9/10. Note, however, that these numbers can only be considered approximate since ambiguities with assumptions of slightly altered non-turbulent velocity laws within the nebulae cannot be excluded \citep{sabbadin.08}. Thus, shares of 0 and 1 resp.~are also conceivable. Moderate turbulence (even up to the speed of sound of about 10 \kms) changes the shape of the line profiles as tested by, e.g., \citet{gesicki.03} and \citet{MoSt.08}, but hardly the peak seperation itself. Of course this holds only as long as the turbulence is of Gaussian distribution, and the orginal peak seperation is not much smaller than the assumed turbulent velocity component.

Since our study concentrates on PNe with H-rich central stars, we expect neither the shape of the outer wings of the line profiles, which typically are the high-velocity signature of the post-shock region of the shell, nor the peak-seperation of the inner components significantly affected by turbulence.}

\section{On the evaluation of the true nebular expansion velocities}\label{eval.veloc}

In the previous section we found convincing evidence, from theory as well as from observations, that lines of N$^+$ are to be preferred for tracing the expansion of PNe.  We arrived then at a mean (spectroscopic) expansion velocity close to 30~\kms\ that is a bit higher than values used in the literature. We repeat her, however, that this spectroscopically derived value refers to the flow property of bulk material with the highest emission measures along the lines-of-sight and is not to be taken if one is concerned about expansion ages and visibility times. Then one must derive the true expansion velocity, which is given by the propagation of the outer shock, $\dot{R}_{\rm out}$.
   
\begin{figure}
\vskip-2mm
\hskip-3mm
\includegraphics*[width=1.03\linewidth]{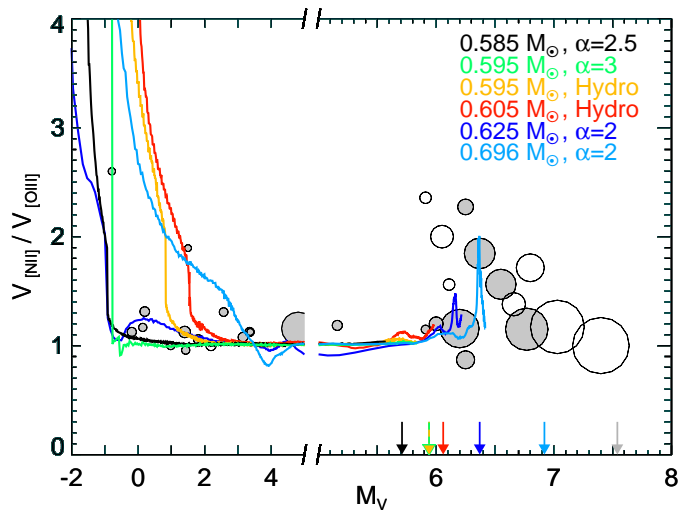}
\caption{The predicted variations of ${F = V_{\rm \nii}/V_{\rm \oiii}}$ with stellar absolute
         magnitude are compared with measurements of the 
         $V_{\rm \nii}$ and $V_{\rm \oiii}$ velocities listed in Table \ref{basic.data}.  
         The abscissa is expanded for ${M_V \ge 5}$ for more 
         clarity.  The model sequences used are indicted in the legend.  Evolution proceeds from
         left to right, and the
         limiting stellar magnitudes reached after 25\,000 years of evolution are indicated along 
         the abscissa by the (coloured) vertical arrows.  For comparison, the grey vertical arrow at 
         ${M_V = 7.55}$ belongs to a 0.94 \Msun\ post-AGB model. 
         The size of the symbols corresponds to the nebular radius: from the smallest objects,
         0.03~pc (\object{NGC 7027}, no. 46, and \object{NGC 6567}, no. 146) to the largest, 1.09 pc 
         (\object{PuWe\,1}, no. 5). 
         Smaller objects are a bit enlarged for a better visibility, but the size relations are
         kept qualitatively.   Open symbols have
         distances ${D\le 1}$ kpc, filled symbols have ${D > 1}$~kpc.  The typical error of 
         the stellar brightness is, due to the distance uncertainty, about $\pm$0.5~dex.
        }
\label{moe.2a}
\end{figure}

We showed in the previous section that often velocities measured from \nii\ are higher than those from \oiii.  This property is now illustrated in Fig.~\ref{moe.2a} in more detail.  In this figure we have plotted the ratio ${F = V_{\rm \nii}/V_{\rm \oiii}}$ over stellar absolute magnitude $M_V$ as predicted by our nebular models.  These model predictions are compared with observed velocity ratios as taken from Table~\ref{basic.data}. 
    
During the early phase with a rather cool, luminous central star we see a remarkable difference between $V_{\rm \nii}$ and $V_{\rm \oiii}$ which is due to the nebular ionisation stratification: N$^+$ lines traces the faster moving outer parts of the shell, while O$^{2+}$ resides preferentially in the innermost parts which had been decelerated in the past by the (thermal) gas pressure.  The ratio can be quite high for a short time, depending on the properties of the model sequence.  The object with the highest observed ratio in this stage is \object{IC~418} (no. 76) with ${F = 2.60}$ at $M_V = -0.79$.

If the nebula becomes optically thin, $F$ rapidly drops to unity because both ions extend smoothly over the whole nebula, albeit with different concentrations.  This happens at quite different stellar magnitudes and depends on the relation between the time scales of stellar evolution and nebular expansion.  Note that the models of the 0.696 \Msun\ sequence never become optically thin.  Yet, $F$ approaches unity at high stellar temperatures because, next to N$^+$, O$^{2+}$ also becomes a minority species.
  
For low-luminosity central stars, recombination establishes again an ionisation stratification and causes $F$ to increase up to about two for some time before $F$ decreases slowly again with increasing reionisation.   Only models with a stellar core of 0.696 \Msun\ reach here temporarily a value of $F$ as high as $\approx$2 (cf. Fig.~\ref{moe.2a}), because here the ionisation stratification is predominant.

Our models are consistent with the observations: They cover the observed trend of $F$ with stellar evolution, and velocity ratios higher than unity occur only for bright, young and faint central stars. In both stages PNe are optically thick (or partially thick) with significant ionisation stratification and with a positive velocity gradient. None of the PNe with faint central stars has $F$ significantly higher than two (Fig.~\ref{moe.2a}).  
 
We note here that our sequences with the massive central stars (0.625 and 0.696 \Msun) do reach the observed low stellar magnitudes, but their ages (and the radii of their nebular models) are still very small (cf. left panels of Fig.~\ref{moe1}). Since the stellar luminosity and temperature does not change much during the remaining part of evolution, and assuming that also the density gradient ahead of the nebular edge remains roughly constant during the remaining part of nebuler expansion, we expect that  $V_{\rm \nii}/V_{\rm \oiii}$ will not change appreciably and will stay close to unity for the rest of evolution.
   
The limiting stellar magnitudes $M_V$ of those massive central star models at ages of 25\,000$\pm$5000 years are 6.38$\pm$0.07 (0.625 \Msun) and 6.92$\pm$0.06 (0.696 \Msun).  The two largest PNe seen in Fig.~\ref{moe.2a} have also the faintest nuclei (${M_V \geq 7}$) and are \object{A\,31} (no. 9, ${M_V = 7.0}$) and \object{PuWe\,1} (no.~5, ${M_V = 7.4}$). Provided that the distances are correct, both PNe must harbour an extremely massive central star:  The 0.94 \Msun\ post-AGB models of \cite{B.95}, e.g., reach $M_V = 7.55\pm0.06$ mag after 25\,000$\pm$5000 years of post-AGB evolution.

We conclude that it is impossible to derive individual corrections for PNe with unknown $V_{\rm \nii}$ since the evolutionary stage of an object in question is not sufficiently known.  For statistical purposes, however, the following averages of spectroscopic expansion velocities for objects with faint central stars (${M_V \ge 5}$) of the two samples ($\le$1 kpc and $>$1 kpc) and where both N$^+$ and O$^{2+}$ lines are available may be useful:
\begin{itemize}
  \item  $D\le 1$ kpc (7 PNe): \\
         $\langle V_{\rm \nii}\rangle\, \, = 33.4\pm6.8$ \kms, \\
         $\langle V_{\rm \oiii}\rangle = 22.7\pm8.4$ \kms, and
  \item  $D> 1$ kpc (9 PNe): \\  
         $\langle V_{\rm \nii}\rangle\,\,  = 27.4\pm5.4$ \kms, \\
         $\langle V_{\rm \oiii}\rangle = 21.2\pm7.1$ \kms,      
\end{itemize}
with a mean ratio ${\langle F \rangle = 1.5\pm0.5}$ (all 16 objects).  Thus, the typical observed spectroscopic expansion velocity of a PN belonging to the subample with faint central stars is about 30 \kms, with a dispersion of 6 \kms, based on the \nii\ lines. 

Based on the previous results we worked out the following route for deriving true expansion velocities for our local PNe sample:
\begin{enumerate}   
  \item  First of all, we decided to determine the propagation speed of the physical radius
         of PNe, corresponding to $R_{\rm out}$ of our models.  The use of the 10\,\% 
         isophote radius is too ambiguous because the latter depends on
         density and ionisation structure and can rapidly change during
         evolution (cf. Fig.~\ref{r10.vs.rout}).
  \item  Select $V_{\rm \nii}$ based on the peak separation of split emission 
         lines.  If \nii\ is not available, lines from O$^+$ or S$^{2+}$ ars also good
         choices.
  \item  The main task is now to convert a measured gas velocity into the true expansion
         velocity of the PN outer edge by means of our radiation-hydodynamic 
         models, i.e. to determine a (theoretical) correction factor $F_{\rm N\,II}$ 
         such that $\dot{R}_{\rm out} = F_{\rm N\,II}\times V_{\rm \nii}$. 
\end{enumerate}

\begin{figure}
\vskip-0.1cm
\includegraphics*[width=\linewidth]{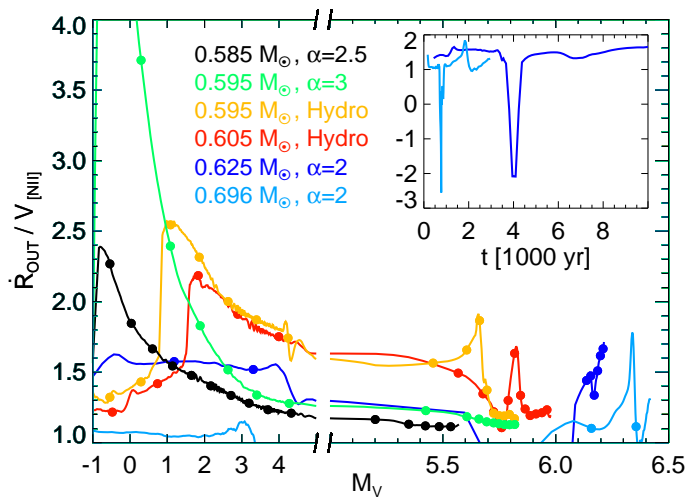}
\caption{Correction factor $F_{\rm N\,II}$ (=\,${\dot{R}_{\rm out}/V_{\rm \nii}}$)
         as it follow from various radiation-hydrodynamic
         model sequences, plotted over stellar absolute magnitude, $M_V$. 
         Evolution is always from left to right.  The abscissa is broken at 
         ${M_V = 5}$, and the faint part is expanded to provide more clarity.
         As compared to Fig.~\ref{moe1}, three additional sequences were included. 
         The dots along the velocity sequences correspond to time steps of 10$^3$ years each
         and are meant only to illustrate the pace of evolution, not absolute post-AGB ages. 
         The 0.595 \Msun\ sequence with ${\alpha = 3}$
         reaches very early a maximum  $F_{\rm N\,II} ({= \dot{R}_{\rm out}/V_{\rm \nii}})$ of 7.5.              
         The inset illustrates the time evolution of $F_{\rm N\,II}$
         during the nebular recombination phase (0.625 and 0.696 \Msun\ only) when
         $F_{\rm N\,II}$ becomes smaller than unity and even negative for a very brief time span.
         Note that for the earliest evolutionary stages, ${M_V \le -1}$, our predictions
         can be uncertain because of the extremely small sizes of the model nebulae.
         }
\label{moe.2b}
\end{figure}

The relation between the spectroscopic ``expansion" velocities based on nitrogen, $V_{\rm \nii}$, and the true expansion given by $\dot{R}_{\rm out}$, expressed by a correction factor $F_{\rm N\,II}$, is presented in  Fig.~\ref{moe.2b}.  We see that $F_{\rm N\,II}$ is very model dependent, but the trend is the following: During the early, optically thick phase (small $M_V$), the correction factor is quite small, between about 1.1 and 1.4. At this stage, nitrogen traces matter closely behind the (D-type) ionisation front. During the transion to the optically thin stage this correction factor $F_{\rm N\,II}$ ``jumps" up to much  higher values (to 7.5 in the case of the 0.595 \Msun, ${\alpha = 3.0}$ sequence) because the line peaks come from the denser wind-compressed rims which expand much more slowly (see, e.g., \citetalias{schoenetal.05a}; \citealt{jacobetal.12}). The size of this jump is larger for the sequences with steep density gradients because the shock propagates then faster.

During the following evolution the corrections decrease until the star fades because the rim acceleration is higher than that of the outer shock.  Responsible is the ever increasing thermal pressure of the wind-shock heated bubble gas.

An exception is the 0.696 \Msun\ sequence because the model nebulae never become optically thin during the high-luminosity part of the central-star evolution.  Therefore, the correction factor remains close to one until recombination sets in at stellar magnitude ${M_V \approx 3.5}$.  During the nebular recombination phase, the shock is decelerated because the electron temperature behind the shock drops, but the nebula matter continues to expand at the same rate.  Consequently, $F_{\rm shock}$ becomes even negative until the post-shock matter is heated again due to re-ionisation ($M_V \approx 6$).  The inset reveals, however, that this phase is extremly short: 190 yr for the 0.696~\Msun\ case and 430 yr for the 0.625~\Msun\ case.  
  
In principle, for an anticipated expansion time of some 10\,000 years, the mass-loss history along the final AGB evolution should be always considered.  This, however, would demand a very time comsuming modelling, as we did for the ``hydro'' cases.  Of special importance are the thermal pulses which lead to a short but strong variation of the AGB mass loss. Nevertheless, we neglect these complications because (i) these mass-loss variations are very short, (ii) $F_{\rm N\,II}$ appears to stay above unity, as our ``hydro'' models suggest, and (iii) the thermal pulse frequency is quite low as compared to the PN expansion times, viz. about 100\,000 years for a remnant mass of 0.6 \Msun\ and about 40\,000 years for 0.7 \Msun\ \citep[e.g.,][]{Pa.70, schoen.79, WZ.81}.

We conclude from Fig.~\ref{moe.2b} that it is difficult to find an appropriate correction factor $F_{\rm N\,II}$ for PNe with luminous central star ($M_V < 5$) because it varies strongly with stellar brightness and mass, and also with ionisation status. Instead we used a method introduced successfully by \citet{corradetal.07} to directly measure the post-shock velocity. The hydrodynamical models used in this study suggest that the ratio between shock and post-shock velocity, $F_{\rm post}$, is small and rather independent of the evolutionary state: ${\dot{R}_{\rm out}= F_{\rm post}{\times}  V_{\rm post} = (1.2\ldots 1.3){\times} V_{\rm post}}$ \citep[see Fig.~5 in][]{corradetal.07}. This holds for the optically thin phase of evolution only. For an early optically thick model, $F_{\rm post} \approx 1$.  

\begin{figure}
\hskip-1mm
\includegraphics*[width=1.01\linewidth]{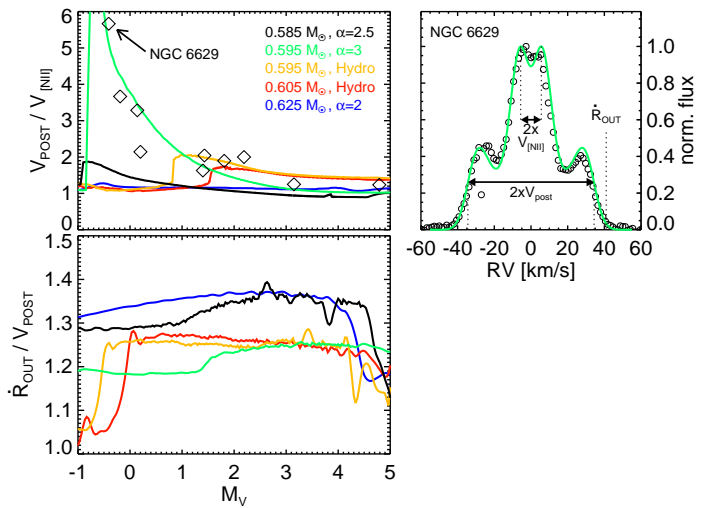}
\caption{Evolution of corrections factors. \emph{Top left panel}: the theoretically predicted variations of 
         $V_{\rm post}/ V_{\rm \nii}$ with absolute stellar magnitude, 
         $M_V$, are compared  with observationally determined values of $V_{\rm post}$ 
         and $V_{\rm \nii}$ for PNe for which both velocities are known 
         (see Table~\ref{basic.data}).
         The maximum value of $V_{\rm post}/ V_{\rm \nii}$ is 7.1 right after
         the thick/thin transition of the 0.595 \Msun\ ``${\alpha = 3}$'' model
         sequence.
         \emph{Bottom panel}:  here the ratio ${F_{\rm post} = \dot R_{\rm out}/V_{\rm post}}$ 
         vs. absolute stellar magnitude is plotted for the same simulations shown in the
         {top left panel}.
         \emph{Top right}: observed line profile of \nii\ $\lambda$\,6583 \AA\ (circles) of
         the PN \object{NGC 6629} (no. 196), taken from Paper\,II compared to a theoretical 
         profile computed from a model of the 0.595~\Msun\ ''${\alpha = 3}$" sequence, convolved with
         a Gaussian of 10~\kms\ FWHM.  This model corresponds closely to the
         position of \object{NGC 6629} in the top left panel, and its parameters are the
         following:  $\teff = 46\,800$~K, (post-AGB) age = 3968 yr, and ${M_V = -0.58}$, 
         ${V_{\rm \nii} = 5.6}$~\kms, and ${V_{\rm post} = 34.4}$~\kms.  Observed values are
         ${\teff = 47\,000}$~K \citep{mendezetal.88}, ${M_V = -0.41}$,  ${V_{\rm \nii} = 6}$~\kms, 
         and ${V_{\rm post} = 34}$~\kms\ (see Table~\ref{basic.data}, no. 196). 
         The vertical dotted  line indicates the shock velocity of the model, 
         $\dot{R}_{\rm out} = 41$ \kms. 
        }
\label{post.shock}
\end{figure}

Figure \ref{post.shock} illustrates the situation for the hydrodynamical sequences used here. In the top panel we have plotted the predicted ratio $V_{\rm post}/ V_{\rm \nii}$ versus stellar absolutes magnitude.  The individual curves are much smoother than those of Fig.~\ref{moe.2b}, simply because now two gas velocities are compared.  The general trend, however, is the same:  $V_{\rm \nii}$ which traces the rim material increases faster with time than $V_{\rm post}$ does. The velocity difference can be quite  high shortly after the transition to optically thin models: e.g. about 30~\kms\ for the 0.595 \Msun\ ``${\alpha = 3}$'' model sequence, with a ratio $V_{\rm post}/ V_{\rm \nii} = 7.1$\footnote{The ``jumps'' from ratios close to 1 towards higher values for every sequence are caused by a characteristic change of the emission-line profile shapes: at early phases the line profiles of low-ionised species, such as N$^{+}$, are often quad-peaked \citepalias[cf.][Fig.~3 therein]{schoenetal.05b} depending on spectral resolution.  Due to ionisation stratification the outer shell-components are stronger than the inner rim-components at first. However, as $V_{\rm \nii}$ we strictly measured the peak separation between the maximum components which will become the inner components only in time. The exact times these jumps occur depend complicatedly on initial and boundary conditions of each model run (e.g. absolute densities, abundance distributions, SEDs) and are beyond to be adjusted with the limited grid of sequences we employ in this study. Had we chosen the inner of the \nii\ components, which mainly originate in the rim, the jumps would have occured earlier and would have been even larger.}. 
    
The local PN sample listed in Table \ref{basic.data} contains 10 PNe for which, next to $V_{\rm \nii}$, also the post-shock velocity, $V_{\rm post}$, is available from \citet{jacobetal.12} or \citet{corradetal.07}, and these are also shown in the top panel of Fig.~\ref{post.shock}.\footnote{The measurements of $V_{\rm post}$ in \cite{corradetal.07} are based on the \oiii\ 5007 \AA\ line, but the post-shock velocity is, by virtue of the method of its measurement, independent of the ion used, provided the nebula is optically thin.} There is a smooth decline of the observed ratio $V_{\rm post}/ V_{\rm \nii}$ with $M_V$, with a very gratifying agreement with the model predictions.  This indicates a close correspondence, on the average, of our geometrically rather simple models with the expansion properties of real objects.  The best correspondence with reality shows the ``${\alpha = 3}$'' sequence with the 0.595 \Msun\ central star whose nebular models are also the fastest expanding ones, in terms of $\dot R_{\rm out}$. The object with the brightest nucleus has the highest velocity ratio (${V_{\rm post}/ V_{\rm \nii} = 5.7}$) and is \object{NGC 6629} (No. 196), that with the faintest nucleus has the lowest ratio (${V_{\rm post}/ V_{\rm \nii} = 1.2}$) and is \object{NGC 2610} (No. 155).
   
The predictive power of our models is further validated by the example shown in the top right panel of Fig.~\ref{post.shock}.  There we show the \nii\ $\lambda$\,6583\,\AA\ emission line profile of \object{NGC 6629} and compare it with an appropriate theoretical line profile. We emphasise that this sequence was not explicitly designed to fit this object!  The relevant model and PN parameters are given in the figure caption.  Here we have a good example of how far away the peak-separation velocities can be from the real expansion speed.  The latter amounts to 41~\kms, as compared to $V_{\rm \nii} = 6$~\kms!
 
This sequence of decreasing ${V_{\rm post}/ V_{\rm \nii}}$ with decreasing stellar brightness (or increasing age) is excellent evidence for the typical internal kinematics of a PN as outlined in the Introduction. While the leading edge/shock at $r= R_{\rm out}$ propagates with moderate acceleration, only ruled by the density slope and the electron temperature, the inner rim is steadily accelerated by the increasing power of the central-star wind; from close to stalling at the beginning of evolution up to a speed which is later on comparable to that of the edge. 

The bottom panel of Fig.~\ref{post.shock} presents the model-predicted correction factors $F_{\rm post} =  \dot R_{\rm out}/V_{\rm post}$ along the horizontal part of evolution in the \changed{HRD}.  Contrary to $F_{\rm N\,II}$, $F_{\rm post}$ does not vary much during evolution and also not from model to model sequence:  from about 1.2 to about 1.35, with a tendency for the lower values if the    (initial) circumstellar density slope is steeper!  Only during the early optically thick stage, and later during recombination (if it occurs at $M_V \rightarrow 5$), $F_{\rm post}$ may come closer to unity.  
    
We note that  $F_{\rm post}$ depends on the physical properties of the flow, e,g. on the shock strength, which then explains the difference in $F_{\rm post}$ between the model sequences and also the slight variation with evolution. Considering that obviously PNe expand into circumstellar matter with a rather steep density gradient, ${\alpha > 2}$ (e.g. \citealt{sandin.08}; \citetalias{schoenetal.05a}), a value of $F_{\rm post} = 1.25\pm0.05$ appears appropriate.

\section{Estimation of visibility times}\label{vis.time}

Based on the assumption that in a volume-limited sample of PNe all objects up to a certain radius can be observed, visibility times can be easily derived once the spectroscopical expansion speeds are converted into real ones and a meaningful average expansion velocity is found.  Here we assume, like in \citet{frew.08} and \citet{MoDe.06}, that ${R_{\rm PN} = 0.9}$ pc is as reasonable radius limit for the local sample. In accordance with the previous section we only consider objects with spectroscopically measured velocities that are exclusively based on the peak separation of \nii\ lines or with known post-shock velocities.

Because of the transion time of the central star from the tip of the AGB until it is able to ionise the circumstellar matter and to create a PN, the minimum observed nebular radius is certainly larger than zero.  However, the choice of an appropriate minimum initial PN radius is not critical since we have conservatively estimated in Sect.~\ref{models} that the expected initial nebular radii are certainly not larger than 0.05~pc. The smallest nebular radius in our sample, 0.02~pc, belongs to \object{M\,1-26} (No. 78) with a very bright central star, ${M_V \approx -1}$, and low spectroscopic expansion  velocity, $V_{\rm \nii}= 6$ \kms\ (Tab.~\ref{basic.data}). Considering distance uncertainties and the dispersion of observed expansion velocities, this small offset from zero of the initial nebular radii can safely be neglected. 

Likewise, we assume in the following that the observed nebular radii, $R_{\rm PN}$, correspond to the true physical radius.  Thus we neglect the small difference between the H$\alpha$-10\,\% isophote radius and the true radius, given by the position of the leading shock.  The discussion of our models in Sect.~\ref{radius} suggest that this is a very reasonable assumption.  We have also seen from our models that for most of the nebular evolution $\dot{R}_{10} \approx \dot{R}_{\rm out}$ holds, the only exception being rapid phases of recombination and/or density variations of the circumstellar matter. Thus, with the correction factors derived from the hydrodynamic models, we get the true PN expansion velocity, $\dot{R}_{\rm PN}$, by ${\dot{R}_{\rm PN} = F_{\rm N\,II}\times V_{\rm \nii}}$ or by ${\dot{R}_{\rm PN} = F_{\rm post}\times V_{\rm post}}$.

\begin{figure}
\includegraphics[width=1.07\linewidth]{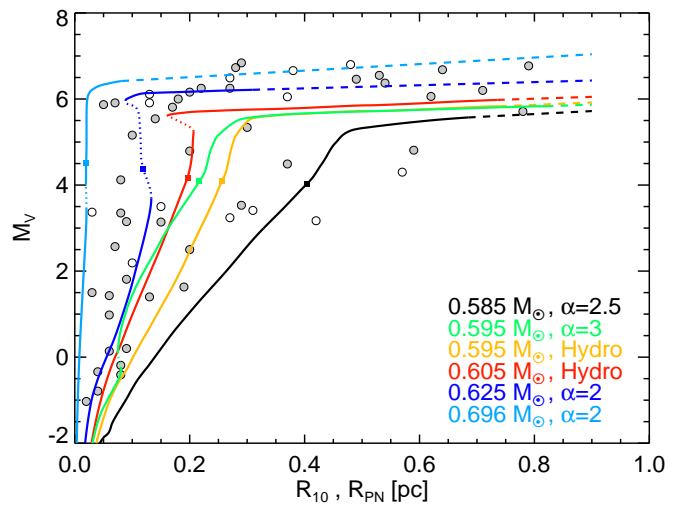}
\caption{Observed and computed
         stellar absolute magnitudes, $M_V$, vs. observed/modelled radii, $R_{\rm PN}$, 
         $R_{10}$.  The model sequences shown are indicated in the legend, and the
         symbols refer to all objects from Table \ref{basic.data}, again broken up into
         local objects (distance $\leq$1 kpc, open) and those with distance $>$1 kpc (filled).
         Evolution goes from the lower left corner towards the upper right corner.
         The dotted parts of the model tracks (0.696, 0.625, and 0.605 \Msun\ ``hydro'' only) 
         indicate strong recombination with a reduction of $R_{10}$ (cf. discussion 
         in Sect.~\ref{radius}).
         The different slopes of the model sequences correspond mainly to the speed of stellar 
         evolution, modified by the nebular expansion properties. 
         The coloured squares mark the position of maximum stellar temperature and hence
         divide the sequences into the high-luminosity and low-luminosity parts.
         Dashed lines indicate that the evolution has been extrapolated assuming conservatively
         that the shock propagation of the last computed model remains constant.        
        }
\label{MV.vs.radius}
\end{figure}

In order to distinguish the observed objects and to put them into certain mass categories of their central stars, a diagram where the stellar (absolute) brightness, $M_V$, is plotted against the nebular radius $R_{\rm PN}$, overlaid with predictions of our model sequences (plotted over $R_{10}$), is quite useful (Fig.~\ref{MV.vs.radius}). Although the detailed run of stellar brightness and nebular radius depends somewhat on the nebula's expansion properties (cf. the two 0.595 \Msun\ sequences with different initial envelope assumptions), the spread of the theoretical sequences is mainly due to the mass-dependent speed of post-AGB evolution, possibly modified by recombination which may temporarily reduce the observable radius, $R_{10}$: The most massive central stars spent only very little time along the high-luminosity part of evolution, while the least massive central stars remain for a very appreciable time at high luminosities.  Considering time spans of several 10\,000 years, even a PN with a central star of moderate mass, say of 0.6 \Msun, spends most of its life in the low-luminosity reionisation phase.  
   
We note a surprisingly good agreement between the observations and our models. Considering the uncertainty of observed stellar magnitudes (at least $\pm0.75$ due to the distance calibration only), a few of the faintest objects may have nuclei heavier than 0.63 \Msun, but 0.7 \Msun\ seems to be a real upper mass limit (cf. Fig.~\ref{moe3})!   We recall that a 0.7 \Msun\ post-AGB model becomes as faint as $M_V \approx 6.9$ mag after 20\,000 years of evolution, and none of the faint objects from Frew's local sample shown in Fig.~\ref{MV.vs.radius} appears fainter than about 7 mag. On the low-mass side very few objects seem to have a stellar core with less than 0.585 \Msun.
   
The comparison of the observations with our model sequences suggests then the following: (i) objects with bright nuclei and not too small radius belong to the class with less massive central stars (${{\cal M}\la 0.63}$ \Msun), and (ii) objects with faint nuclei (${M \ga 5}$ mag) are certainly a mixture of objects with all possible stellar masses.

\subsection{PNe with low-luminosity central stars}

We additionally divide the subsample of PNe with low-luminosity central star into objects with $R_{\rm PN} < 0.15$ pc (4 PNe) and $R_{\rm PN} \geq 0.15$ pc (15 PNe).   The former should exclusively consist of PNe with massive nuclei, the latter of PNe with both massive and less massive central stars.  The mean velocities for both groups are \\[3pt]
$\langle V_{\rm \nii}\rangle = 33.8\pm 3.6$ \kms\ ($R_{\rm PN} < 0.15$ pc) and \\[3pt]
$\langle V_{\rm \nii}\rangle = 29.5\pm 6.5$ \kms\ ($R_{\rm PN} \geq 0.15$ pc), \\[3pt]
i.e. they are, within the dispersions, equal.   

The correction factor, $F_{\rm N\,II}$, has been discussed in connection with Fig.~\ref{moe.2b}: it is (for ${M_V > 5}$) about 1.3 in the 0.696~\Msun\ case and about 1.6 in the 0.625 \Msun\ case. Note that the recombination stage where this correction falls below 1 is too short to be considered.  ${F_{\rm N\,II}\approx 1.2}$ is appropriate for those objects with less massive central stars which do not recombine much, if at all.

Since central stars with a mass of 0.7 \Msun\ are certainly rarer than those with 0.63~\Msun, we chose ${F_{\rm N\,II} = 1.5}$ for all PNe with ${R_{\rm PN} < 0.15}$~pc, leading to a mean real expansion speed\footnote{We note that the uncertainties associated with the choice of the correction factor, $F_{\rm N\,II}$ is only of about 10\,\%, much smaller than the width of the observed velocity distribution.  We thus have neglected it in the error budget .} of this sample of ${\langle\dot{R}_{\rm PN}\rangle = 1.5{\times}33.8\,\mathrm{km\,s^{-1}} = 50.7\pm 5.4}$ \kms. The average time spent in this radius range is then ${2900\pm 300}$ years.  For the mixed group with $R_{\rm PN} \geq 0.15$\,pc we chose $F_{\rm N\,II} = 1.3$ as a weighted mean between 1.2 and 1.6 and get ${\langle\dot{R}_{\rm PN}\rangle = 1.3{\times}29.5\,\mathrm{km\,s^{-1}} = 38.4\pm 8.5}$ \kms.  The corresponding time spent is then ${(0.9-0.15)\,\mathrm{pc}/\langle\dot{R}_{\rm PN}\rangle = 19\,100\pm 4300}$ yr.  The previous high-luminosity part of their evolution occurs with higher expansion rates (see next subsection), and we have ${0.15\,\mathrm{pc}/(48\pm9)\,\mathrm{km\,s^{-1}} = 3100\pm 600}$ yr, adding up to a total of 22\,200\,$\pm$\,4400 years.
   
Assuming that for the PNe with massive nuclei (${{\cal M} > 0.63}$ \Msun) only the faint stage really counts (Fig.~\ref{MV.vs.radius}), we still have a visibility time (${0.9\,\mathrm{pc}/38.4\,\mathrm{km\,s^{-1}}}$) of only 22\,900 years, with a dispersion of 5100 years.  This is probably an upper limit, provided these PNe with massive central stars continue to expand faster, as found for the $<$0.15-pc sample.

\subsection{PNe with luminous central stars}    

We consider here all objects with ${M_V < 5.0}$ (and ${D< 2}$ kpc) from Table \ref{basic.data} to belong to this group, altogether 18 PNe with known $V_{\rm \nii}$.  

As we have seen in the pevious section, the conversion from  $V_{\rm \nii}$ to $\dot R_{\rm out}$ can be quite considerably and changes rapidly with the evolutionary stage, also heavily depending on the model structure used (cf. Fig.~\ref{moe.2b}). We thus preferred instead to only use objects with measured  post-shock velocity, $V_{\rm post}$. This reduces the number of available objects to only 10 (from 18), but the great advantage is that we know the correction factor rather precisely from our models: ${F_{\rm post} = 1.25}$ over the whole considerd stellar brightness range.

\begin{table}
\caption{\label{bright} Real expansion velocities of PNe, averaged over absolute brightness intervals 
         of their central stars.
        }
\centering
\tabcolsep=7.8pt
\begin{tabular}{r c c c c}
\hline\hline\noalign{\smallskip}
 $M_V$-Rang\rlap{e}  & $\langle V_{\rm post}\rangle$ & Corr.  & $\langle\dot{R}_{\rm PN}\rangle$ & No. of  \\[1.5pt]
 [mag]~~             & [\kms]                        & Factor & [\kms]                           & Objects \\[1.5pt] 
\hline\noalign{\smallskip}
 $-1\ldots 1$        & $33.8\pm9.0$                  & 1.25   & $42\pm11$                        & 4       \\
 $1\ldots 3$         & $41.8\pm4.8$                  & 1.25   & $52\pm\phantom{0}6$              & 4       \\
 $3\ldots 5$         & $38.5\pm5.0$                  & 1.25   & $48\pm\phantom{0}6$              & 2       \\[3pt]
 $>5$                & --                            & --     & $39\pm\phantom{0}9$              & 19\enspace \\[1.5pt]
\hline
\end{tabular}
\end{table}

The expansion of the outer shock of these PNe is known to accelerate to some extent (\citetalias[cf.][]{schoenetal.05a}; \citealt{jacobetal.12}) due to a steep circumstellar/halo density gradient and increasing electron temperature within the nebula, thus binning over stellar brightness intervals appears to be necessary. Table \ref{bright} lists the results for the three ${\Delta M_V = 2}$ bins, ranging from $-1\ldots 5$, hence covering nearly all of the horizontal evolution across the \changed{HRD} towards maximum stellar temperature.  There are too few objects as to detect a significant acceleration with progress of evolution, thus we consider the \changed{plain average ${\langle\dot{R}_{\rm PN}\rangle = 47\pm 9}$~\kms\ as a typical real expansion velocity} of double-shell PNe with central-star masses of about 0.6 \Msun\ in this particular evolutionary phase.\footnote{\changed{The \citet{jacobetal.12} sample of PNe with measured post-shock velocities contains altogether 22 PNe because no distance limit is invoked.  The mean post-shock velocity of this larger sample is $\langle V_{\rm post}\rangle = 36.4 \pm 6.1$ \kms, hence ${\langle\dot{R}_{\rm PN}\rangle = 46\pm 8}$~\kms}, i.e. virtually identical with the result drawn from the subsample used here.} This value is higher than our 0.605 \Msun\ ``hydro'' sequence predicts, $\langle\dot{R}_{\rm out}\rangle = 30.6$ \kms, for ${M_V<5}$, but very similar to the expansion property of our 0.595 \Msun, ``${\alpha = 3}$" sequence, $\langle\dot{R}_{\rm out}\rangle = 50.2$ \kms, for $M_V<5$ (cf. also top panels of Fig.~\ref{moe1}).
     
A post-AGB star of typically 0.6 \Msun\ remains about 8000 years at the luminous part of the \changed{HRD} ($M_V \la 5$), hence the nebular radius will grow up to \changed{about 8000 yr\,$\times$\,47 \kms\ $\approx$ 0.39 pc}.  For the rest (up to ${R_{\rm PN} = 0.9}$ pc), the evolution is a bit slower, 0.51 pc / 39~\kms\ = 12\,800 yr.   In total, a lifetime of about 21\,000 years follows then, with an uncertainty of about $\pm$5000 years. Less massive central stars (which are more abundant) evolve more slowly, hence they spent more of the total PN lifetime at the luminous stage with relatively high nebular expansion velocities.  Hence the nebular visibility time of these objects will be even smaller.
 
For comparison, the bottom row of Table \ref{bright} reports our result for the ensemble of PNe with faint ($M_V > 5$) central stars but with $R_{\rm PN} \ge 0.15$~pc. One may indeed interpret the difference as indication of a slight deceleration of the nebular expansion once the central source has faded, which is, again, in line with the majority of our models (see, again, top panels of Fig.~\ref{moe1}).

\subsection{The visibility time}

The rather simple excercises in the previous subsection demonstrate that the total lifetime of a PN is composed of contributions from the two main evolutionary phases, i.e. the phase with a bright and the phase with a faint central star.  The relative importance of these two phases depend on the speed of evolution across the HRD, hence on the stellar mass. In order to get a typical visibility time of the whole PN ensemble of a stellar system, one should evaluate mean values of the real expansion velocities for the different evolutionary stages and compute then a weighted average over the central-star mass distribution.  This task is practically impossible to perform since, even provided the spectroscopic velocities are known, the evolutionary stages and the central-star masses are poorly known or not known at all.
   
Fortunately we have seen from the local sample investigated here that observed expansion velocities do not seem to vary much with evolution and/or central-star mass.  Also, the observed sample is presently too small to allow a division into different even smaller subsamples. Thus we continue with the derivation of a meaningful average of the real expansion velocity as follows:  We just average over all stellar magnitudes (including ${M_V<5}$) and arrive at ${\langle\dot{R}_{\rm PN}\rangle = 42\pm 10 }$ \kms, which then yields a mean visibility time (as defined here) of 0.9\,pc/42\,\kms\ = 21\,000$\pm$5000 years.

In conclusion from all the results obtained above \textit{we consider a visibility time of 21\,000$\pm$5000 years for all PNe with solar/Galactic disk composition as a reasonable value for statistical studies.} We emphasise that this time is entirely determined by the expansion property of the nebula, it is only weakly dependent on the mass of the central star and its speed of evolution across the \changed{HRD} and the following decline in luminosity.

\subsection{Metallicity and visibility times}

So far we have implicitly assumed that all PNe studied are metal-rich, with a elemental distribution typical for the Galactic disk/solar neighbourhood. Likewise, our model sequences used here refer to this particular chemical composition, named by us $Z_{\rm GD}$ \citepalias[see, e.g.,][]{perinotto.04}.  The metallicity influences the expansion behaviour via the electron temperature:  the latter is higher for metal-poorer objects, and hence their expansion is faster than for objects with high metallicities. This effect has extensively been studied recently in \citetalias{schoenetal.10}. 

In order to get an impression of the effect of metallicity on the PN visibility times, we compared four hydrodynamical sequences, all with the same central star model and the same initial envelope configuration: 0.605 \Msun\ nucleus and initial ``hydro" radial density distribution.  The metallicity is varying from $3Z_{\rm GD}$, \ldots, $Z_{\rm GD}/10$ in steps of 0.5 dex.  
 
\begin{table}
\caption{\label{metal.vis} Expansion of a hydrodynamic nebular model with different metallicities
         but otherwise with the same parameters: 0.605~\Msun\ central star and initial ``hydro" 
         envelope. 
        }
\centering
\tabcolsep=11.5pt         
\begin{tabular}{lccc}
\hline\hline\noalign{\smallskip}
         Metallicity     &  Post-AGB Age  &  $R_{\rm out}$  &  $\langle\dot{R}_{\rm out}\rangle$ \\[1.5pt]
                         &  [yr]          &  [pc]           &  [\kms]                            \\[1.5pt]
\hline\noalign{\smallskip}   
         $3Z_{\rm GD}$   &  20\,000       &  0.67           &  32.8                              \\
 \enspace$Z_{\rm GD}$    &  20\,000       &  0.74           &  36.2                              \\
 \enspace$Z_{\rm GD}/3$  &  20\,000       &  0.87           &  42.6                              \\
 \enspace$Z_{\rm GD}/10$ &  20\,000       &  0.94           &  46.1                              \\[1.5pt]
\hline             
\end{tabular}        
\end{table}

Some relevant numbers are collected in Table~\ref{metal.vis}.  It shows the sizes of the different nebular models after 20\,000 years of hydrodynamical simulation, and also the mean expansion velocities, as function of the metal content.  These velocity differences are due to the metal-dependent line cooling of the gas:  less cooling means higher electron temperatures and faster expansion. Table~\ref{metal.vis} shows, however, that the influence of metallicity on the visibility times is modest:  A metallicity change by --1 dex leads to a corresponding decrease in visibility time of about 30\,\% only (--0.15 dex).

\section{Discussion and conclusion}\label{discuss}

\subsection{Real expansion velocities} 

We succeeded in determining real expansion velocities and new visibility times based on a sample of round/elliptical Milky Way PNe with well-known spectroscopically measured expansion velocities. We found that, using information provided by sophisticated 1D radiation-hydrodynamics models, the typical expansion speed of a PN, averaged over the whole lifetime, is about 40 \kms.
 
This quite substantial increase in the canonical mean expansion speed from 20 or 25 \kms\ to a value of 40 \kms\ is due to two facts not considered before:
\begin{itemize}
  \item  Exclusively using velocities measured from \nii\ lines (or from equivalent ions) yields 
         faster expansion than lines from \oiii:  about 27 \kms\ from \nii\ instead of about 
         22 \kms\ from \oiii.\footnote{Only objects with both velocities measured.} 
         If we consider more evolved objects of, say, at least 0.15\,pc radius 
         (cf. left panels of the last two rows in Fig.~\ref{moe1}) the mean velocities are very 
         similar, 30 \kms\ from \nii\ and 23 \kms\ from \oiii.       
  \item  The hydrodynamical modelling revealed that the peak separations of emission 
         lines generally do not trace the fast expanding outer parts of a PN.
         Guided by the hydrodynamical models we are able to derive corrections which allow us to estimate
         the true expansion which is, on the average, about 40~\kms\ instead, i.e. nearly twice as fast
         as one gets from \oiii\ lines.  
\end{itemize}  
  
While the determination of the true nebular expansion velocities is based on a welth of observational material which only needs to be corrected in an appropiate way, the influence of the nebular metallicity on the expansion behaviour (and visibility time) is entirely based on our models:  They predict a slight increase/decrease in the expansion/visibility time with decreasing metallicities. 
  
Our study, of course, only refers to nebulae with well-defined radii, i.e. bipolar objects are omitted.   Their expansion velocity is expected to vary strongly with distance from the equator, hence a meaningful mean value is difficult to estimate.  Fortunately, their fraction to the whole PN population is small, probably not excceeding about 10\,\%, as we stated already in Sect.~\ref{SB.radius}. For instance, the local sample of PNe with measured expansion velocities only contains five extremely bipolar objects. \secchanged{Note, however, that censuses of this kind only consider shapes as they appear in the plane of the sky and do not take possible projection effects into account. Regarding bipolars seen pole-on and thus misclassified, \cite{phillips.01} estimated from statistics on observed aspect ratios for the collimation disks in 25 bipolar nebulae that the true number of bipolars could be a factor 1.7 higher than usually assumed -- still a minority of objects.}

Because we had to resort to \nii\ (or \oii) lines only, the number of useful PNe is rather small since most spectroscopic work has concentrated, for obvious reasons, on the strong \oiii\ lines. More observational efforts to measure spectroscopic expansion velocities by means of \nii\ lines (or from other singly ionised ions) would be very welcome, in addition to measurements of the post-shock velocities as in \cite{corradetal.07} and \cite{jacobetal.12}, or gross shell velocities as in \citetalias{schoenetal.05a}.  We are, however, convinced that the present PN sample yields already reliable results which are very important for further studies.
         
The local sample selected by \citet{frew.08} is also contaminated by objects with a hydrogen-deficient central star.  They appear to expand moderately faster than the rest of the sample (cf. Table~\ref{Wolf-Rayet}).  We found an average value of about $35\pm7$ \kms\ for the spectroscopic expansion velocity, instead of about 30 \kms\ for objects with a normal central star. Also typical is that ${V_{\rm \oiii} \ge V_{\rm \nii}}$, except during the reionisation phase (e.g. A\,21, no. 16 in Table~\ref{Wolf-Rayet}). We see here obviously the effect of the very powerful winds emanating from hydrogen-deficient stars of Wolf-Rayet spectral type which compress and accelerate the inner nebular regions much stronger than is the case for stars with a normal surface composition.
 
Though we do not have appropriate hydrodynamical models at hand, it is sure that the real expansion velocity is correspondingly higher than the spectroscopically measured value. Assuming a correction factor of 1.3, a real average expansion with 45~\kms\ follows for nebulae hosting hydrogen-deficient nuclei.

The PNe with bona fide binary central stars listed in Table~\ref{binaries} seem to have exceedingly low spectroscopic expansion velocities, i.e. not more than 25 \kms.  Their number is, however, too small as to allow firm conclusions.  But taken at face value, one may conclude that PN formation dominated by tidal interaction leads to a different expansion behaviour than for (apparently) single stars for which photoionisation and wind interaction dominate.
   
Finally we note that only a rather small fraction ($\approx$18\,\%) of the more than 200 objects in Frew's \citeyearpar{frew.08} local PNe sample were useful for our velocity study, i.e those for which $V_{\rm \nii}$ from a split line profile is avalaible.   It would be important to increase this fraction considerably, also for allowing improvement of the statistics on sub groups of objects (bipolar, binary core, hydrogen-deficient core, \ldots) which are not well, or not at all, represented in this study.

\subsection{Visibility times} 
  
The visibility time, or total lifetime of a PN up to a size of ${R_{\rm PN} = 0.9}$~pc, as discussed here and in the literature, is exclusively ruled by the nebula's real expansion velocity whose run with age depends on the circumstellar environment and the evolution of the electron temperature.  The mass, i.e. the evolutionary speed, of the central star is only of minor importance. Indeed, we also found from the existing observations that the mean nebular expansion does not appreciably depend on stellar mass.  Thus it is justified to use for a volume-limited sample of PNe with maximum nebular radius of 0.9\,pc only one typical value for the PN visibility time, which is, as we have shown, 21\,000$\pm$5000 years.
  
The situation is different in extragalactic work:  There we have to deal with flux-limited PN samples, and the ``visibility time" depends on the depth of the respective survey. This is illustrated in Fig.\,15 (lower panel) of \citet[][ hereafter Paper IV]{schoenetal.07} which displays the brightness evolution of model nebulae belonging to sequences with different central stars and initial envelope configurations (some of them also used in the present work). For instance, if the depth of the survey extends to only two magnitudes below the bright PN cut-off at $M(5007) = -4.5$ mag \citep{ciard.03}, only PNe with stellar cores more massive than about 0.59 \Msun\ can be detected because only they are able to generate a nebula sufficiently bright in $M(5007)$. The corresponding lifetimes  within this magnitude range is only up to about 5000 years or less, depending on the core mass.  In order to get a visibility time of 20\,000 years, the survey must go as deep as seven magnitudes below the PN cut-off brightness!

\subsection{The PN death rate and the birthrate of white dwarfs}   
 
The higher mean nebular expansion velocity as found in this work has, of course, direct implications for the birth-/death rate of PNe since it scales directly with expansion velocity. Since the central stars of PNe evolve directly into white dwarfs (WDs), one expects a close correspondence between the local death rate density of PNe and the local birthrate density of WDs.  The latter has been determined empirically by \cite{liebert.05} as
\begin{displaymath}
       \chi_{\rm WD} =  (1.00\pm0.25)\! \times\! 10^{-12}  \rm\quad pc^{-3}\,yr^{-1} .
\end{displaymath}
\cite{MoDe.06} predicted by their population synthesis a local formation rate density of WDs (from single and binary stars) of 
\begin{displaymath}
       \chi_{\rm WD} =  (1.1\pm0.5)\!\times\! 10^{-12} \rm\quad pc^{-3}\,yr^{-1}. 
\end{displaymath}   
Using a surface density of local PNe\footnote{All morphological types are considered.} with radii $<$0.9 pc of ${\mu = 13\pm 1}$ kpc$^{-2}$ as determined by \citet{frew.08} together with a scale height $h$ of $220\pm25$ pc, we get a local PN space density \citep[][ Eq. 11.3 therein]{frew.08} of
\begin{displaymath}
       \rho_{\rm PN} = 0.5\, \mu / h = 30 \pm 5 \rm\quad kpc^{-3} . 
\end{displaymath}   
 The corresponding local PN death rate density follows then from
\begin{displaymath}
       \chi_{\rm PN} = \rho_{\rm PN}\!\times\! \dot{R}_{\rm PN} / \Delta R_{\rm PN} 
\end{displaymath}           
\begin{displaymath}
       \phantom{\chi_{\rm PN} }=   (1.4\pm0.5)\! \times\! 10^{-12}  \rm\quad pc^{-3}\,yr^{-1} ,
\end{displaymath}               
where we have used ${\dot{R}_{\rm PN} = 40\pm15}$ \kms,  $\Delta R_{\rm PN} = 0.9$ pc, and $\rho_{\rm PN} = 30\pm5$ kpc$^{-3}$.  Considering the overlap of the uncertainties, we arrive at a very satisfactory agreement between the local PN death rate density and the local WD birthrate density.
    
However, when comparing death rates of PNe with birthrates of WDs one has to consider that not only PNe are the progenitors of the latter:  \cite{MoDe.06} estimate that only about 70\,\% of WDs are descendants of the PNe phase. The remaining WDs are descendants of stars that cannot climb the AGB to the necessary luminosities, either because the are not massive enough or because of binary action.  The \citeauthor{MoDe.06} estimate is in line with the work of \cite{ds.85} who found that probably only 80\,\% of all WDs are descendents of PNe.  Also \citet{WK.84} estimated that about 30\,\% of WDs could not have been formed via the PN channel.

On the other hand, one has also to consider that the \cite{liebert.05} result still underestimates the true WD formation rate because of undetected faint WDs hidden in binary systems, but this fraction is difficult to estimate and thus still unkown.   

Taken our derived death rate density of PNe and the corresponding birthrate density of WDs in the Galactic disk at face value, we conclude that the former is too large by a factor of about 1.7.  There are, however, a number of caveats:
\begin{itemize}
  \item  First of all, the scale height $h$ to be used for estimating the local PN density is uncertain
         and under discussion.  PNe descent from progenitor stars of quite different masses (and ages)
         whose distribution vertical to the Galactic plane is then also different.
         We preferred a value close that recommended by \cite{frew.08}, but a higher value cannot be 
         excluded and would lead to a lower local PN density, accordingly (see above).  For instance, 
         the measured mean heights above the Galactic plane vary from about 100~pc (bipolar and 
         type\,I PNe) over about 200~pc (elliptical PNe) to about 320~pc (round PNe), according to 
         \citet[Table 11.1 therein]{frew.08}.  A weighted mean is difficult to determine because 
         a complete census of the different morphologies is necessary.
  \item  Another important input for birth-/death rate determination is, next to the completeness
         of the survey and the scale height, the distance scale which enters with the 4th power.  
         Reduction of $\chi_{\rm PN}$ by a factor of 1.7 demands then only a general distance 
         increase by 14\,\%, i.e. of 0.06 dex only, at most.
         Such a small shift is hardly to be seen in Fig.~\ref{moe3} and is within the uncertainties
         of Frew's \citeyearpar{frew.08} distance calibration: A zero-point uncertainty of 10--20\,\%
         is estimated by the author. 
  \item  Also, one has to consider that the local WD birthrate is a value integrated over several 
         gigayears of Milky Way star formation, while the local PN death rate reflects the present 
         death rate of AGB stars in the solar neighbourhood.      
\end{itemize}    
 
We mention in this context that the method of a general scaling of existing PN distances has already a long tradition: By means of plausibility or other more physical arguments, already \citet{Wei.77, Wei.89} showed that the distances derived by \cite{CK.71} are, on the average, too small by a factor of 1.3, at least.  This result is, so to speak, an anticipation of the work done by \cite{frew.08} whose distance calibration provides, on the average, distances that are larger by a factor of 1.35 than those of \citeauthor{CK.71} \citep[cf. Table 7.5 in][]{frew.08}.  The additional distance factor increase proposed here gives then an overall increase in the \cite{CK.71} distances by a factor of 1.5, one of the largest PN distance scales up to date.   
    
We emphasise, however, that a slight general increase in the \citet{frew.08} distances as proposed here is still consistent (within the errors) with the few existing trigonometric parallax measurements. \changed{\cite{benedict.09} determined HST parallaxes for three of the apparently closest PNe of our sample,  \object{NGC\,7293} (no. 2), \object{NGC\,6853} (no. 6), and \object{A\,31} (no. 9), that lie within ranges of 105--93\%, 114--100\%, and 148--114\%\, of the nominal distances adopted by \citet{frew.08}.}

We note in this context that \citeauthor{frew.08} applied his distance calibration on PNe of the Magellanic Clouds and found a slight underestimation of the distances to  both Clouds by about --4\,\%. Correcting this, the Frew calibration contains then a scale factor of 1.4 against the old \cite{CK.71} calibration, which is only 7\,\% smaller than the one derived here. 
    
In any case, it is clear from the above discussion that the WD formation rate is much better constrained than the death rate of PNe with its possibly quite large systematic uncertainties.  For the latter, it is the scale height and/or the scale factor (or both together) that possibly conspire against a more precisely limited formation-/death rate of PNe. Thus, given  all these uncertainties, and also considering the different completeness distances, 10~pc vs. 1000 pc, the close agreement between the local PN death rate and the WD birthrate is, in our opinion, quite astonishing. We think that too detailed interpretations of possible differences between these two rates and comparisons with predictions from population synthesis work should be considered with utmost caution.

\begin{acknowledgements}
We thank  Max Moe for extensive discussions subsequent to IAU Symp. 283 on the importance of the visibility time for statistical studies of planetary nebula samples in, e.g., the context of population synthesis studies. 
\end{acknowledgements}

\noindent \emph{Note added in proof.}\enspace
The recently published paper of Pereyra et al.~(2013, ApJ 771, A2) reported a deceleration of nebular expansion in evolved PNe. Peak-separation velocity data of PNe which are in common in both studies agree for the most part. We note that their sample is not volume-limited. The 2-kpc sample of our Table~\ref{basic.data} could be increased by 8 (out of 100) objects from Pereyra et al.~(2013). Unfortunately, combining the velocity information of both studies yields only two additional PNe for which then both \nii\ and \oiii\ velocities are available, viz. Sh 2-200 and K 1-22. Thus, the numbers presented in our paper would only change in the decimals. 
  
We have also concluded that a small decrease in PN expansion rates of evolved PNe may occur, but we have also stressed that peak-separation velocities are not adequate to determine true expansion rates. Nevertheless, if we perform a plain statistical analysis for the entire volume-limited sample of Table~\ref{basic.data} supplemented by the data from Pereyra et al.~(${D \la 2}$ kpc), we still hardly see any deceleration of peak-seperation velocities (cf. Fig.~\ref{moe1}): ${\langle V_{\rm \oiii}\rangle}$ decreases only slightly from $28.8 \pm 9.4$~\kms\ for the 15 PNe with ${3 \le M_V < 5}$ to $25.1 \pm 10.2$~\kms\ for ${M_V \ge 5}$ (28 PNe). For nitrogen emission lines we see no deceleration at all, ${\langle V_{\rm \nii}\rangle} = 27.2 \pm 7.1 $~\kms\ (${3 \le M_V < 5}$, 5 PNe) vs. ${\langle V_{\rm \nii}\rangle} = 28.7 \pm 7.9 $~\kms\ (${M_V \ge 5}$, 24 PNe). We adhere to our argument that such decelerations are either only apparent due to ionisation stratification or, if indeed real, only a transient phenomenon: flow speeds can be reduced during a possible phase of recombination, but quickly recover former expansion rates once re-ionisation sets in. Regarding the hint of Pereyra et al.~(2013) that experiencing a possible interaction with the interstellar matter (ISM) might be another source for PN deceleration, it seems worthwhile to correlate velocity studies such as our and Pereyra et al.~with the statistical studies on PNe--ISM interaction, e.g., Ali et al.\ (2012, A\&A 541, A98).



\appendix

\section{Details of the PNe sample used}\label{appendix1}

\begin{table*}
\caption{\label{basic.data}
         List of all objects used in the present study with their relevant properties.  
        }
\tabcolsep=3.0pt
\begin{tabular}{rp{2.5cm}ccccccccl}
\hline\hline\noalign{\smallskip}
  No.  & Name    & $D$  & $R_{\rm PN}$ & $\log S_{\rm H\alpha}$                & $M_V$ & ${V_{\rm exp}}$ & $V_{\rm \nii}$ & $V_{\rm \oiii}$ & ${V_{\rm post}}$ & Remarks \\[1.5pt]
       &         & [kpc]& [pc]         & [erg\,cm$^{-2}$\,s$^{-1}$\,sr$^{-1}$] & [mag] & [\kms]          & [\kms]         & [\kms]          & [\kms]           & \phantom{\citetalias{schoenetal.05a}, H07, JSLZSS12} \\[1.5pt]  
  (1)~ &\ \, (2) & (3)  & (4)          & (5)                                   & (6)   & (7)             & (8)            & (9)             & (10)             & \quad (11) \\[1.5pt]
\hline\noalign{\smallskip} 
    2   &  NGC 7293   &  0.22  &  0.48  &       --4.02  &         6.80  &  21\rlap{\tablefootmark{a}}  &  24       &  14       & \ldots  &          \\  
    5   &  PuWe 1     &  0.37  &  1.09  &       --5.53  &         7.40  &  23\rlap{\tablefootmark{a}}  &  27       &  27       & \ldots  &          \\ 
    6   &  NGC 6853   &  0.38  &  0.37  &       --3.42  &         6.05  &  32\rlap{\tablefootmark{a}}  &  30       &  15       & \ldots  &          \\
    7   &  NGC 1360   &  0.38  &  0.31  &       --4.06  &         3.41  &  34\rlap{\tablefootmark{a}}  &  \ldots   &  28       & \ldots  &          \\  
    8   &  A 36       &  0.45  &  0.42  &       --4.91  &         3.17  &  36                          &  \ldots   &  35       & \ldots  &          \\  
    9   &  A 31       &  0.48  &  1.18  &       --5.31  &         6.99  &  29\rlap{\tablefootmark{a}}  &  35       &  \ldots   & \ldots  & HW90     \\
   20   &  EGB 6      &  0.59  &  1.03  &       --5.73  &         7.03  &  25\rlap{\tablefootmark{a}}  &  41       &  35       & \ldots  &          \\
   27   &  Sh 2-200   &  0.66  &  0.57  &       --4.83  &         4.30  &  13                          &  \ldots   &  13       & \ldots  &          \\    
   28   &  NGC 6720   &  0.70  &  0.13  &       --2.41  &         6.11  &  22\rlap{\tablefootmark{a}}  &  39       &  25       & \ldots  & ODSH07   \\    
   29   &  NGC 7008   &  0.70  &  0.15  &       --2.97  &         3.50  &  37                          &  \ldots   &  39       & \ldots  &          \\
   33   &  NGC 3587   &  0.76  &  0.38  &       --3.88  &         6.66  &  34\rlap{\tablefootmark{a}}  &  40       &  37       & \ldots  & GCMK03   \\
   39   &  NGC 3132   &  0.81  &  0.13  &       --2.66  &         5.91  &  21                          &  33       &  14       & \ldots  & H07      \\  
   44   &  IC 5148/50 &  0.85  &  0.27  &       --3.94  &         6.49  &  53                          &  \ldots   &  53       & \ldots  &          \\
   46   &  NGC 7027   &  0.89  &  0.03  &\enspace 0.07  &         3.37  &  22                          &  27       &  24       & \ldots  &          \\ 
   51   &  NGC 4361   &  0.95  &  0.27  &       --3.47  &         3.24  &  32                          &  \ldots   &  26       & \ldots  &          \\
   53   &  NGC 3242   &  1.00  &  0.10  &       --1.78  &         2.19  &  28                          &  20       &  20       &  40     & \citetalias{schoenetal.05a}\\ 
   58   &  A 34       &  1.03  &  0.71  &       --5.36  &         6.20  &  35                          &  37       &  32       & \ldots  &          \\    
   62   &  NGC 6153   &  1.10  &  0.07  &       --1.15  &         2.57  &  17                          &  17       &  13       & \ldots  &          \\
   70   &  A 33       &  1.16  &  0.78  &       --5.11  &         5.71  &  32                          &  \ldots   &  32       & \ldots  &          \\    
   75   &  NGC 6772   &  1.20  &  0.22  &       --3.02  &         6.25  &  11                          &  25       &  11       & \ldots  &          \\
   76   &  IC 418     &  1.20  &  0.04  &       --0.27  &\llap{--}0.79  &  16\rlap{\tablefootmark{a}}  &  13       &\enspace 5 & \ldots  & \citetalias{schoenetal.05a}\\  
   78   &  M 1-26     &  1.20  &  0.02  &\enspace 0.01  &\llap{--}1.03  &  12                          &\enspace 6 &  \ldots   & \ldots  &          \\
   80   &  IC 1295    &  1.23  &  0.30  &       --3.68  &         5.34  &  27                          &  34       &  \ldots   & \ldots  & GVM98    \\   
   81   &  NGC 7662   &  1.26  &  0.09  &       --1.60  &         3.15  &  27\rlap{\tablefootmark{a}}  &  28       &  26       &  35     & CSSJ07, Tautenbg.\\
   86   &  NGC 6826   &  1.30  &  0.08  &       --1.42  &\llap{--}0.19  &  16                          &\enspace 9 &\enspace 8 &  33     & JSLZSS12, Tautenbg.\\
   91   &  NGC 6894   &  1.31  &  0.17  &       --2.71  &         5.81  &  43                          &  \ldots   &  43       & \ldots  &          \\
   92   &  A 61       &  1.31  &  0.64  &       --5.20  &         6.68  &  30                          &  \ldots   &  32       & \ldots  &          \\  
   96   &  K 1-22     &  1.33  &  0.62  &       --4.61  &         6.06  &  28                          &  \ldots   &  28       & \ldots  &          \\
   98   &  NGC 2899   &  1.37  &  0.37  &       --3.39  &         4.49  &  25                          &  \ldots   &  32       & \ldots  &          \\     
   101  &  A 62       &  1.38  &  0.54  &       --4.44  &         6.37  &  15                          &  24       &  13       & \ldots  &          \\
   102  &  NGC 6445   &  1.39  &  0.14  &       --1.94  &         5.54  &  38                          &  \ldots   &  38       & \ldots  &          \\
   106  &  A 39       &  1.40  &  0.59  &       --5.09  &         4.81  &  29                          &  37       &  32       & \ldots  & HW90     \\       
   107  &  Wray 17-31 &  1.40  &  0.49  &       --4.31  &         6.46  &  28                          &  28       &  \ldots   & \ldots  &          \\
   111  &  NGC 2438   &  1.42  &  0.27  &       --3.37  &         6.25  &  23                          &  20       &  23       & \ldots  &          \\ 
   114  &  NGC 7009   &  1.45  &  0.09  &       --1.26  &         1.81  &  25\rlap{\tablefootmark{a}}  &  19       &  18       &  36     & JSLZSS12, Tautenbg.\\ 
   117  &  NGC 6804   &  1.47  &  0.19  &       --2.78  &         1.63  &  25                          &  \ldots   &  25       & \ldots  &          \\
   120  &NGC 7076/A 75&  1.47  &  0.20  &       --3.51  &         2.50  &  42                          &  \ldots   &  42       & \ldots  &          \\     
   127  &  NGC 6543   &  1.50  &  0.09  &       --1.11  &         0.20  &  19\rlap{\tablefootmark{a}}  &  21       &  16       &  45     & JSLZSS12, Tautenbg.\\  
   130  &  IC 2149    &  1.52  &  0.04  &       --1.00  &\llap{--}0.34  &  20                          &  20       &  \ldots   & \ldots  &          \\   
   135  &  IC 4593    &  1.57  &  0.06  &       --1.67  &         0.14  &  12                          &\enspace 7\rlap{\tablefootmark{a}} &\enspace 6\rlap{\tablefootmark{a}} &  23  & JSLZSS12, Tautenbg.\\ 
   136  &  NGC 6905   &  1.58  &  0.15  &       --2.72  &         3.14  &  40                          &  \ldots   &  44       & \ldots  &          \\
   140  &  NGC 7354   &  1.60  &  0.13  &       --1.31  &         1.40  &  25                          &  27       &  24       &  44     & JSLZSS12, Tautenbg.\\   
   142  &  NGC 6818   &  1.64  &  0.10  &       --1.74  &         5.16  &  27                          &  32       &  27       & \ldots  &          \\
   145  &  NGC 6563   &  1.67  &  0.20  &       --2.90  &         6.16  &  11                          &  \ldots   &  11       & \ldots  &          \\
   146  &  NGC 6567   &  1.68  &  0.03  &       --0.65  &         1.50  &  18                          &  36       &  19       & \ldots  &          \\  
   147  &  A 84       &  1.68  &  0.53  &       --4.42  &         6.55  &  16                          &  25       &  16       & \ldots  &          \\
   151  &  NGC 5882   &  1.70  &  0.06  &       --1.08  &         1.43  &  \ldots                      &  23       &  24       &  47     & \citetalias{schoenetal.05a}, H07, JSLZSS12\\   
   155  &  NGC 2610   &  1.72  &  0.20  &       --3.53  &         4.79  &  24                          &  \ldots   &  34       &  42     & \citetalias{schoenetal.05a}, CSSJ07\\ 
   158  &  NGC 4071   &  1.77  &  0.28  &       --3.40  &         6.73  &  17                          &  \ldots   &  17       & \ldots  &          \\
   159  &  A 51       &  1.78  &  0.29  &       --4.05  &         3.53  &  42                          &  \ldots   &  42       & \ldots  &          \\
   169  &  NGC 3918   &  1.84  &  0.08  &       --1.01  &         3.35  &  25                          &  27       &  24       & \ldots  &          \\  
   173  &  NGC 6572   &  1.86  &  0.06  &       --0.58  &         0.98  &  18                          &  16       &  16       & \ldots  &          \\
   180  &  NGC 3211   &  1.91  &  0.07  &       --1.97  &         5.91  &  27                          &  31       &  27       & \ldots  & $V_{\rm \oii}$ \\  
   183  &  A 13       &  1.93  &  0.79  &       --4.51  &         6.77  &  20                          &  23       &  20       & \ldots  &          \\
   189  &  NGC 7048   &  1.97  &  0.29  &       --3.44  &         6.84  &  15                          &  \ldots   &  17       & \ldots  &          \\    
   195  &  NGC 6537   &  2.00  &  0.05  &       --0.46  &         5.87  &  18                          &  \ldots   &  18       & \ldots  &          \\
   196  &  NGC 6629   &  2.00  &  0.08  &       --1.29  &\llap{--}0.41  &  \enspace 6                  &\enspace 6 &  \ldots   &  34     & \citetalias{schoenetal.05a}, CSSJ07\\  
\hline
\end{tabular}
\end{table*}

\addtocounter{table}{-1}
\begin{table*}
\caption{continued.
         }
\tabcolsep=3.0pt
\begin{tabular}{rp{2.5cm}ccccccccl}
\hline\hline\noalign{\smallskip}
  No.  & Name    & $D$  & $R_{\rm PN}$ & $\log S_{\rm H\alpha}$                & $M_V$ & ${V_{\rm exp}}$ & $V_{\rm \nii}$ & $V_{\rm \oiii}$ & ${V_{\rm post}}$ & Remarks \\[1.5pt]
       &         & [kpc]& [pc]         & [erg\,cm$^{-2}$\,s$^{-1}$\,sr$^{-1}$] & [mag] & [\kms]          & [\kms]         & [\kms]          & [\kms]           & \phantom{\citetalias{schoenetal.05a}, H07, JSLZSS12} \\[1.5pt]  
  (1)~ &\ \, (2) & (3)  & (4)          & (5)                                   & (6)   & (7)             & (8)            & (9)             & (10)             & \quad (11) \\[1.5pt]
\hline\noalign{\smallskip} 
   199  &  NGC 3195  &  2.01  &  0.18  &  --2.69  &         6.00  &          26  &        30 &   25      & \ldots  & $V_{\rm \oii}$ \\    
   202  &  NGC 2792  &  2.02  &  0.08  &  --1.92  &         4.12  &          20  &    \ldots &   20      & \ldots  &                   \\
\hline 
\end{tabular}
\tablefoot{The entries in Cols. 1--6 correspond to those in Tables 9.5 and 9.6 of \citet{frew.08}. 
           Column 7 gives the velocities measured and compiled respectively by \citet[][his Tables 3.8\tablefootmark{a} and 9.4]{frew.08}. 
           Columns 8 and 9 contain the velocities based on line peak separations 
           as retrieved from the literature; they are from \citet{W.89} if not otherwise noted. 
           Column 10 lists the post-shock velocities measured by \citet{corradetal.07} or
           \citet{jacobetal.12}.  Column 11 gives references/remarks referring to the 
           velocities listed in Cols. 8 and 9.\\
           \tablefoottext{a}{inner components of emission line profile not split; velocity from 
            half width at half maximum of the central Gauss profile.}
          }
\tablebib{CSSJ07: ${V_{\rm post}}$ from \citet{corradetal.07};
          H07: $V_{\rm \nii}$ ($V_{\rm \oiii}$) from \citet{H07};
          HW90: \citet{HW90}; 
          GCMK03: \citet{GCMK03};
          GVM98: \citet{GVM98};
          JSLZSS12: ${V_{\rm post}}$ from \citet{jacobetal.12};
          ODSH07: \citet{ODSH07};   
          Tautenbg.: $V_{\rm \nii}$ and/or $V_{\rm \oiii}$ from \'echelle spectrograms taken at Tautenburg Observatory, Germany (Sch\"onberner
          et al., in prep.).  
         }
\end{table*}

\begin{table*}
\caption{\label{Wolf-Rayet}
         Same as in Table \ref{basic.data} but for objects with hydrogen-deficient nuclei only.}
\tabcolsep=3.0pt
\begin{tabular}{rp{2.5cm}ccccccccl}
\hline\hline\noalign{\smallskip}
  No.  & Name    & $D$  & $R_{\rm PN}$ & $\log S_{\rm H\alpha}$                & $M_V$ & ${V_{\rm exp}}$ & $V_{\rm \nii}$ & $V_{\rm \oiii}$ & ${V_{\rm post}}$ & Remarks \\[1.5pt]
       &         & [kpc]& [pc]         & [erg\,cm$^{-2}$\,s$^{-1}$\,sr$^{-1}$] & [mag] & [\kms]          & [\kms]         & [\kms]          & [\kms]           & \phantom{\citetalias{schoenetal.05a}, H07, JSLZSS12} \\[1.5pt]  
  (1)~ &\ \, (2) & (3)  & (4)          & (5)                                   & (6)   & (7)             & (8)            & (9)             & (10)             & \quad (11) \\[1.5pt]
\hline\noalign{\smallskip} 
   10   &  NGC 246         &  0.50  &  0.29  &          --4.11  &         3.30  &  35\rlap{\tablefootmark{a}}  &  \ldots   &  40       &  \ldots &  OVI/PG 1159 \\  
   16   &  A 21/YM 29      &  0.54  &  0.89  &          --4.72  &         7.21  &  32\rlap{\tablefootmark{a}}  &  45       &  \ldots   &  \ldots &  PG 1159     \\ 
   30   &  NGC 1501        &  0.72  &  0.09  &          --2.49  &         3.13  &  40                          &  40       &  40       &  \ldots &  WC4, NAGS00 \\  
   48   &  Jn 1            &  0.90  &  0.74  &          --4.95  &         6.11  &  36\rlap{\tablefootmark{a}}  &  \ldots   &  15       &  \ldots &  PG 1159     \\
   57   &  NGC 40          &  1.02  &  0.11  &          --1.95  &         0.20  &  25                          &  25       &  30       &  \ldots &  WC8, NAGS00 \\
   69   &  JnEr 1          &  1.15  &  1.06  &          --4.98  &         6.77  &  24                          &  41       &  22       &  \ldots &  PG 1159     \\
   88   &  BD 30\degr 3639 &  1.30  &  0.02  &\phantom{--}0.10  &\llap{--}1.07  &  25                          &  28       &  36       &  \ldots &  WC9, BM99   \\  
  105   &  NGC 7094        &  1.39  &  0.34  &          --4.43  &         2.65  &  40                          &  \ldots   &  45       &  \ldots &  PG 1159     \\
  109   &  NGC 2371-72     &  1.41  &  0.13  &          --2.90  &         3.95  &  43                          &  \ldots   &  43       &  \ldots &  WCE         \\
  112   &  NGC 5189        &  1.44  &  0.46  &          --3.10  &         3.63  &  25                          &  25       &  37       &  \ldots &  WO1         \\
  134   &  NGC 6369        &  1.58  &  0.15  &          --2.72  &\llap{--}0.01  &  42                          &  37       &  36       &  \ldots &  WO3. MPMS06  \\ 
  197   &  NGC 6751        &  2.00  &  0.11  &          --2.23  &         1.60  &  36                          &  38       &  40       &  \ldots &  WO4, Tautenbg.\\ 
  203   &  A 30            &  2.02  &  0.62  &          --5.25  &         2.76  &  40                          &  \ldots   &  40       &  \ldots &  WC, born-again \\      
 \hline 
\end{tabular}
\tablebib{BM99: \citet{BM99}; MPMS06: \citet{MPMS06}; NAGS00: \citet{NAGS00};
          Tautenbg.: $V_{\rm \nii}$ and/or $V_{\rm \oiii}$ from \'echelle spectrograms taken at Tautenburg Observatory, Germany (Sch\"onberner
          et al., in prep.).  
         }
\end{table*}

\begin{table*}
\caption{\label{binaries}
         Same as in Table \ref{basic.data} but for objects which are known to harbour a close binary.}
\tabcolsep=3.0pt
\begin{tabular}{rp{2.5cm}ccccccccl}
\hline\hline\noalign{\smallskip}
  No.  & Name    & $D$  & $R_{\rm PN}$ & $\log S_{\rm H\alpha}$                & $M_V$ & ${V_{\rm exp}}$ & $V_{\rm \nii}$ & $V_{\rm \oiii}$ & ${V_{\rm post}}$ & Remarks \\[1.5pt]
       &         & [kpc]& [pc]         & [erg\,cm$^{-2}$\,s$^{-1}$\,sr$^{-1}$] & [mag] & [\kms]          & [\kms]         & [\kms]          & [\kms]           & \phantom{\citetalias{schoenetal.05a}, H07, JSLZSS12} \\[1.5pt]  
  (1)~ &\ \, (2) & (3)  & (4)          & (5)                                   & (6)   & (7)             & (8)            & (9)             & (10)             & \quad (11) \\[1.5pt]
\hline\noalign{\smallskip}
    4  &  NGC 1514  &  0.37  &  0.12  &  --3.28  &         0.01  & 25         &  \ldots  &  23       & \ldots & MA03\\                   
   21  &  HFG 1     &  0.60  &  0.79  &  --4.80  &         3.06  & 15         &  15   &  13       & \ldots & \\ 
   45  &  NGC 6337  &  0.86  &  0.10  &  --2.47  &         4.13  & \enspace 8 &  \ldots  &\enspace 8 & \ldots & \\  
   47  &  NGC 2346  &  0.90  &  0.14  &  --3.00  &\llap{--}0.08  & 12         &  13   &\enspace 8 & \ldots & \\
   64  &  Sp 1      &  1.13  &  0.20  &  --3.49  &         1.82  & 30         &  \ldots  &  29       & \ldots & \\
   93  &  NGC 6026  &  1.31  &  0.16  &  --3.15  &         1.13  & 25         &  \ldots  &  25       & \ldots & \\     
 \hline 
\end{tabular}
\tablebib{MA03: \citet{muthu.03}.  
         }
\end{table*}

Table \ref{basic.data} contains the detailed information about all the objects considered in the present study.  They contain all objects from the local sample (${D \la 2}$ kpc) selected by \citet{frew.08} and for which velocity information from Doppler split line profiles is available. Strong bipolar objects, and also those with binaries or hydrogen-deficient cores are omitted.  Those are listed in Tables~\ref{Wolf-Rayet} (hydrogen-deficient nuclei) and \ref{binaries} (nuclei in close binaries). The numbering of objects is that of \citet{frew.08}.  In a few cases, we used $V_{\rm \oii}$ instead of $V_{\rm \nii}$ (cf. Col.~11).

Altogether we have 13 objects with a hydrogen-deficient central star listed in Table~\ref{Wolf-Rayet}, i.e. \changed{only 17\,\% of the total local sample with measured spectroscopic expansion velocities used by us (78 PNe). The fraction of known objects with a binary central star is even smaller, only 8\,\% (Table~\ref{binaries}).}

\end{document}